\documentclass[12pt]{article}

\usepackage{amsmath}
\usepackage{amssymb}
\usepackage{amsfonts}
\usepackage{latexsym}
\usepackage{color,subfigure}
\usepackage{graphicx,graphics,epstopdf}
\usepackage{amscd}

\catcode `\@=11 \@addtoreset{equation}{section}

\catcode `\@=12

\newcommand{\be}{\begin{equation}}
\newcommand{\en}{\end{equation}}
\newcommand{\bea}{\begin{eqnarray}}
\newcommand{\ena}{\end{eqnarray}}
\newcommand{\beano}{\begin{eqnarray*}}
\newcommand{\enano}{\end{eqnarray*}}
\newcommand{\bee}{\begin{enumerate}}
\newcommand{\ene}{\end{enumerate}}

\newcommand{\Hf}{{\hat N}_{h}}
\newcommand{\M}{{\hat N}_{m}}

\newcommand{\Mf}{{\hat N}_{m}}
\newcommand{\Cf}{{\hat N}_{s}}

\newcommand{\h}{{h}}
\newcommand{\cf}{{s}}
\newcommand{\hf}{{h}}
\newcommand{\mf}{{m}}

\newcommand{\Sc}{{\cal S}}
\newcommand{\E}{{\cal E}}

\newcommand{\1}{1 \!\! 1}

\newcommand{\Hil}{{\cal H}}

\newcommand{\n}{\vec{n}}

\catcode `\@=11 \@addtoreset{equation}{section}
\catcode `\@=12

\begin{document}
\thispagestyle{empty}
\vspace*{2cm}

\begin{center}
{\Large \bf {Non-hermitian operator modelling  of basic cancer cell dynamics}}   \vspace{2cm}\\

{\large F. Bagarello, F. Gargano}\\
  Dipartimento di Energia, Ingegneria dell'Informazione e Modelli Matematici,\\
Scuola Politecnica, Universit\`a di Palermo,\\ I-90128  Palermo, Italy\\
\end{center}

\begin{abstract}
\noindent We propose a dynamical system of tumor cells proliferation based on operatorial methods. 
The approach we propose is {\em quantum-like}: we use ladder and number operators to describe healthy and tumor cells birth and death,
and the evolution is ruled by a non-hermitian Hamiltonian 
which includes, in a non reversible way,  the basic biological  mechanisms we consider for the system. 
We show that this approach is rather efficient in describing some processes of the cells. We further  add some medical treatment, described by adding a suitable term in the Hamiltonian, which controls and limits the growth of tumor cells, and we propose an optimal approach to stop, and reverse, this growth.
 \end{abstract}

\section{Introduction}
\label{sec:intro}
In the past few years several macroscopic dynamical systems have been discussed using quantum tools. This approach has been considered by many authors and the range of its applications is wide, and has been successfully applied to various fields of sociology and decision making processes \cite{pol1,pol4,all1,all2,qdm5,baggarg,Pkren1,DSO_turncoat2017,DSGO_opinion2017,rev3c,rev3b,BHK,baghavkhr,bbk}, biology  \cite{rev3a,haven2}, ecology \cite{BO_ecomod,BCO2016,DSO}  economics \cite{bag2006,baghav2,Pkren2}, population and crowd dynamics \cite{BO_migration,BGO_crowds,GAR14,therule,BTBB}, etc. More applications of quantum ideas outside a {\em standard} quantum realm can be found, for instance, in the monographs \cite{qal,baa,Busemeyer2012,havkhrebook,Khrennikov2010,bagbook}, and in many of the references cited there.

Even if the underlying idea is shared among several authors (quantum ideas can be used not only in quantum mechanics!),
 our specific approach often differs from that of other authors. In particular, we use ladder operators of different kind (bosonic, fermionic, or generalizations of these) to construct an Hamiltonian operator $H$ for a certain system $\Sc$, and then we use $H$ to deduce the time evolution of the observables of $\Sc$, using the Heisenberg or the Schr\"odinger equations of motion. $H$ is constructed following a set of minimal rules described in \cite{bagbook}, and contains all the interactions occurring between the different {\em agents} of $\Sc$. In most of the cases considered so far, the Hamiltonian $H$  is assumed to be hermitian: $H=H^\dagger$. This allows us to use many standard results of ordinary quantum mechanics. For instance, if an {\em observable} $X$ of $\Sc$, i.e. an hermitian operator acting on  the Hilbert space $\Hil$ where the system is defined, does not depend on the time explicitly, then $X(t)$ is a constant of motion, $X(t)=X(0)$, for all $t\geq0$, if $X$ commutes with $H$. This is because, in the Heisenberg representation, $X(t)=e^{iHt}X(0)e^{-iHt}$, and this coincides with $X(0)$ if $[H,X]=HX-XH=0$.

Sometimes, for some specific application, it is convenient to give up hermiticity of $H$, and to work with some {\em effective Hamiltonian} $H_{eff}$, with $H_{eff}\neq H_{eff}^\dagger$. This is what happens quite often, for instance, in quantum optics and in PT-quantum mechanics, \cite{SZ13,effH1,bagproc,mosta,ben}. In applications to macroscopic systems, as those which are relevant for this paper,  some sort of effective Hamiltonians was introduced, for instance in \cite{BCO2016} and in \cite{BO_ecomod}, to describe stress and positive effects in specific ecological systems. This choice proved to be quite efficient in concrete computations, and is based on the introduction of some imaginary parts in the parameters of the (otherwise hermitian) Hamiltonian of the system. The sign of these imaginary parts is crucial in the modelization procedure. This is not the only possibility to transform an hermitian Hamiltonian into a non-hermitian one. For instance, if $a_1$ and $a_2$ are any two operators relevant in the description of $\Sc$, an hermitian term contributing to $H$ could be $a_1a_2^\dagger+a_2a_1^\dagger$. This, if $a_j$ are ladder operators, represents a {\em reversible} exchange between, say, agent 1 and agent 2, see \cite{bagbook}. If we want to make this exchange  irreversible, the natural choice is to replace the previous sum with a single term, $a_1a_2^\dagger$ for instance: agent 1 is giving something to agent 2, but not viceversa. In this way we clearly loose hermiticity of the Hamiltonian, but we gain in its explicit interpretation. This is exactly the kind of generalization we will adopt here, in the context of cellular division, following and extending the original approach proposed in \cite{haven2}, where ladder operators were also used for a similar problem. In particular we will introduce a manifestly non-hermitian Hamiltonian $H$ describing the formation, the proliferation, and the treatment of a tumor, and we will analyze the time evolution of  some sort of {\em number} of the healthy and sick (tumor) cells. More in details, the basic mechanisms considered in our analysis are the following: (1) an healthy cell becomes sick (because of the presence of some degenerative factor); (2) tumor cells multiply; (3) healthy cells multiply as well, but not at the same rate. {This is because the sick cells, in a realistic tumor evolution, have a faster multiplication dynamics than that of healthy cells; } (4) some treatment for the disease begins. In particular, we will be interested to the effect of the treatment,  to its specific nature, and to the instant in which the treatment begins. Notice that our model is mainly thought to describe some system {\em in vitro}, rather than real patients. For this reason we will restrict our numerical computations to a reasonable, but not too large, number of cells for the system. This is important to keep the computational time under control. { It is also important to stress that in this paper we will not compare the performance of our with those of other existing models. A similar comparison is surely important, but it is not so relevant for us, here; we are more interested in showing that an interesting model can be constructed by using operators, rather than functions, and  quantum-like equations of motion. }

The paper is organized as follows: in the next section we introduce the ladder operators needed in our model. We will focus on some basic facts concerning these operators. They were never used before, in our knowledge, in this kind of applications. In fact, they are neither bosonic nor fermionic, in general, with a dimensionality which is directly connected with the number of cells of our systems. In Section \ref{sec::hamiltonian} we will introduce the Hamiltonian $H$ of the system, and deduce the dynamics out of $H$. Section \ref{sec::numerics} is devoted to our simulations, while our conclusions are contained in Section \ref{sec::conclusions}.

\section{The ladder and the density operators}
\label{sec::ladder}
In many of the applications of quantum ideas to the macroscopic realm, the ladder operators which have been adopted were essentially of two kinds: bosonic or fermionic. This means that our system may have either an infinite number of different conditions (the various eigenstates of the bosonic number operator), or just two\footnote{This is because the fermionic number operator has just two eigenstates.}. These two cases have been enough to discuss many different
applications so far. But, in some cases, it is more convenient to consider some intermediate situation. This is what was done, for instance, in \cite{abjp2013}, in a different context.

Here we show how to construct ladder operators for a finite-dimensional Hilbert space $\Hil_N$, with $N<\infty$. In particular, the interesting case for us is $N\neq 2$, since, when $N= 2$, ladder operators are very well known. For concreteness we will consider in details the construction for $N=5$, and then we will briefly comment on its generalization to other $N$.

Let $\E_5=\{e_j,\,j=0,1,2,3,4\}$ be the canonical orthonormal (o.n.) basis of $\Hil_5$: $\left<e_j,e_l\right>=\delta_{j,l}$, $j,l=0,1,2,3,4$, and $e_0^T=(1,0,0,0,0)$, $e_1^T=(0,1,0,0,0)$ and so on. Here $e_j^T$ is the transpose of $e_j$. We define an operator $b^\dagger$ via its action on the $e_j$'s:
 \be
 b^\dagger e_0=e_1,\quad  b^\dagger e_1=\sqrt{2}\,e_2,\quad b^\dagger e_2=\sqrt{3}\,e_3,\quad b^\dagger e_3=\sqrt{4}\,e_4,\quad b^\dagger e_4=0	
\label{21}\en
We see that $b^\dagger$ behaves as a sort of fermionic raising operator, destroying the upper level. Then, the matrix expression for $b^\dagger$ in the basis $\E_5$ is the following:
$$
b^\dagger=\left(
            \begin{array}{ccccc}
              0 & 0 & 0 & 0 & 0 \\
              1 & 0 & 0 & 0 & 0 \\
              0 & \sqrt{2} & 0 & 0 & 0 \\
              0 & 0 & \sqrt{3} & 0 & 0 \\
              0 & 0 & 0 & \sqrt{4} & 0 \\
            \end{array}
          \right).
$$
Hence its adjoint is
$$
b=\left(
            \begin{array}{ccccc}
              0 & 1 & 0 & 0 & 0 \\
              0 & 0 & \sqrt{2} & 0 & 0 \\
              0 & 0 & 0 & \sqrt{3} & 0 \\
              0 & 0 & 0 & 0 & \sqrt{4} \\
              0 & 0 & 0 & 0 & 0 \\
            \end{array}
          \right).
$$
These operators look like the truncated versions of the raising and lowering bosonic operators. Of course, we cannot expect that they satisfy the canonical commutation relation $[b,b^\dagger]=\1_5$, $\1_5$ being the identity operator in $\Hil_5$, since this would only be possible in an infinite-dimensional Hilbert space. In fact, straightforward computations show that
\be
[b,b^\dagger]=\1_5-5P_4,
\label{22}\en
where $P_4$ is the projection operator on $e_4$: $P_4f=\left<e_4,f\right>\,e_4$, for all $f\in\Hil_5$. Notice that $\1_5-5P_4$ is the following diagonal matrix: $\1_5-5P_4=diag\{1,1,1,1,-4\}$, which differs from the identity matrix on $\Hil_5$ only for the last component in its main diagonal.  $b$ behaves as a lowering operator:
 \be
 b e_0=0,\quad  b e_1=e_0,\quad b e_2=\sqrt{2}\,e_1,\quad b e_3=\sqrt{3}\,e_2,\quad b e_4=\sqrt{4}\,e_3,	
\label{23}\en
as expected. $\hat N=b^\dagger b=diag\{0,1,2,3,4\}$ is the number operator, while $\hat N_s=b\,b^\dagger=diag\{1,2,3,4,0\}$ is a sort of shifted version of $N$, but with a clear difference in the last entry. They satisfy the following eigenvalue equations:
\be
\hat Ne_k=ke_k, \qquad\qquad \hat N_s e_k=\left\{
    \begin{array}{ll}
(k+1)e_k,\quad k=0,1,2,3\\
0,\qquad\qquad\qquad k=4,\\
\end{array}
        \right.
\label{24}\en
$k=0,1,2,3,4$. It is easy to see that $b^5=(b^\dagger)^5=0$.

It is clear how to extend the construction to $N\neq 5$. It is enough to consider the canonical o.n. basis for $\Hil_N$, $\E_N$, and use its vectors to define $b^\dagger$ in analogy with (\ref{21}). Then $b$ is just the adjoint of $b^\dagger$. These are ladder operators such that $b\,e_0=0$ and $b^\dagger e_N=0$, and can be used to define $\hat N$ and $\hat N_s$ as above.

\vspace{2mm}

{\bf Remark:--} If $N=2$ formula (\ref{22}) should be replaced by $[b,b^\dagger]=\1_2-2P_1=diag\{1,-1\}$. Also $b=\left(
                                                                                                       \begin{array}{cc}
                                                                                                         0 & 1 \\
                                                                                                         0 & 0 \\
                                                                                                       \end{array}
                                                                                                     \right)
$ and $b^\dagger=\left(
                                                                                                       \begin{array}{cc}
                                                                                                         0 & 0\\
                                                                                                         1 & 0 \\
                                                                                                       \end{array}
                                                                                                     \right)$, which is in agreement with the well known expressions for the fermionic ladder operators.

\vspace{2mm}

In what follows, we will use this strategy to construct three different families of ladder operators, one for each agent of the biological model we want to describe. Then we will {\em put all these ingredients together}, by taking a suitable tensor product, in order to have a common functional framework. More in details: the agents of the system $\Sc$ are the healthy cells, {\em attached} to the ladder operators $\h$ and $\h^\dagger$, living in an Hilbert space $\Hil_{\h}$. Then we have the sick  cells, described in terms of the ladder operators $\cf$ and $\cf^\dagger$, defined on $\Hil_{\cf}$, and the medical treatment ($\mf$ and $\mf^\dagger$, acting on $\Hil_{\mf}$). We call $N_\alpha=dim(\Hil_\alpha)$, where $\alpha=\h,\cf,\mf$. The o.n. basis  of $\Hil_\alpha$ is $\E_\alpha=\{e_{j}^{(\alpha)}, \, j=0,1,2,\ldots,N_\alpha-1\}$. The operators $\h$, $\cf$ and $\mf$ satisfy relations which extend those above. First of all we have
\be
\h\, e_0^{(\h)}=0,\qquad \cf \,e_0^{(\cf)}=0,\qquad \mf\, e_0^{(\mf)}=0,
\label{25}\en
and then
\be
e_1^{(\h)}=\h^\dagger e_0^{(\h)}, \quad e_2^{(\h)}=\frac{1}{\sqrt{2}}\h^\dagger e_1^{(\h)},\quad\ldots ,\quad e_{N_\h-1}^{(\h)}=\frac{1}{\sqrt{N_\h-1}}\h^\dagger e_{N_\h-2}^{(\h)},
\label{26}
\en
\be
e_1^{(\cf)}=\cf^\dagger e_0^{(\cf)}, \quad e_2^{(\cf)}=\frac{1}{\sqrt{2}}\cf^\dagger e_1^{(\cf)},\quad\ldots ,\quad e_{N_\cf-1}^{(\cf)}=\frac{1}{\sqrt{N_\cf-1}}\cf^\dagger e_{N_\cf-2}^{(\cf)},
\label{27}
\en
and
\be
e_1^{(\mf)}=\mf^\dagger e_0^{(\mf)}, \quad e_2^{(\mf)}=\frac{1}{\sqrt{2}}\mf^\dagger e_1^{(\mf)},\quad\ldots ,\quad e_{N_\mf-1}^{(\mf)}=\frac{1}{\sqrt{N_\mf-1}}g^\dagger e_{N_\mf-2}^{(\mf)}.
\label{28}
\en
Finally, we have that
\be
\h^\dagger e_{N_\h-1}^{(\h)}=0, \quad \cf^\dagger e_{N_\cf-1}^{(\cf)}=0, \quad \mf^\dagger e_{N_\mf-1}^{(\mf)}=0. \quad
\label{29}
\en

The Hilbert space of our system is now the tensor product $\Hil=\Hil_{\h}\otimes\Hil_{\cf}\otimes\Hil_{\mf}$, whose dimension is clearly $N=N_\h\times N_\cf\times N_\mf$. Each operator $X_\h$ on $\Hil_{\h}$ is  identified with the following tensor product on $\Hil$: $X_\h\otimes\1_\cf\otimes\1_\mf$, where $\1_\cf$ and $\1_\mf$ are the identity operators on $\Hil_\cf$ and $\Hil_\mf$, respectively. Analogously, the operators $X_\cf$ and $X_\mf$ on $\Hil_{\cf}$ and $\Hil_{\mf}$ should be identified respectively with $\1_\h\otimes X_\cf\otimes\1_\mf$ and with $\1_\h\otimes \1_\cf\otimes X_\mf$, where we have introduced $\1_\h$, the identity operator on $\Hil_\h$. Furthermore
$$
\left(X_\h\otimes X_\cf\otimes X_\mf\right)\left(f_\h\otimes f_\cf\otimes f_\mf\right)=(X_\h f_\h)\otimes (X_\cf f_\cf)\otimes (X_\mf f_\mf),
$$
for all $f_\h\in\Hil_{\h}$, $f_\cf\in\Hil_{\cf}$ and $f_\mf\in\Hil_{\mf}$. From now on, when no confusion arises, we will just write $X_\h, X_\cf,X_\mf$ instead of $X_\h\otimes\1_\cf\otimes\1_\mf, \1_h\otimes X_\cf\otimes\1_\mf$ and $\1_\h\otimes\1_\cf\otimes X_\mf$, and their action  is obviously intended on the whole $\Hil$.

 An o.n. basis for $\Hil$ is the following:
$$
\E=\left\{\varphi_{n_\h,n_\cf,n_\mf}:=e_{n_\h}^{(\h)}\otimes e_{n_\cf}^{(\cf)}\otimes e_{n_\mf}^{(\mf)}, \, n_\alpha=0,1,\ldots,N_\alpha-1, \, \alpha=\h,\cf,\mf\right\},
$$
{so that any state $\Psi$ of the system $\Sc$ can be expressed as a combination of these vectors:
\bea
\Psi=\sum_{n_\h,n_\cf,n_\mf}c_{n_\h,n_\cf,n_\mf}\varphi_{n_\h,n_\cf,n_\mf}\label{210}.
\ena
Here the sum is extended to all the possible values of $n_\h$, $n_\cf$ and $n_\mf$, and $c_{n_\h,n_\cf,n_\mf}$ are complex scalars not necessarily chosen to normalize  $\Psi$ in the conventional way, $\sum_{n_\h,n_\cf,n_\mf}|c_{n_\h,n_\cf,n_\mf}|^2=1$: we will not necessarily assume that $\|\Psi\|=1$. The reason is that, even if $\|\Psi\|=1$ for $t=0$, it is no longer so, in general, for $t>0$, due to the fact that our time evolution is not unitary, as we will show later.

Each of the elements of $\E$ can be easily interpreted. For instance, $\varphi_{n_\h,0,0},$ with $n_\h>0$, describes a situation in which the system consists only of healthy cells, with no sick cell and with no active medical treatment, whereas  $\varphi_{n,2n,1},$ with $n>0$, represents a state in which the sick cells are twice
the number of the healthy ones, and a medical treatment is acting.
}
A simple computation shows that all the ladder operators of the different agents commute: $[X_s,X_m]=0$, and so on. For instance, using the properties of the tensor product, we see that
$$
[\h, \mf]\varphi_{n_\h,n_\cf,n_\mf}=\h\, e_{n_\h}^{(\h)}\otimes e_{n_\cf}^{(\cf)}\otimes \mf\, e_{n_\mf}^{(\mf)}-\h\, e_{n_\h}^{(\h)}\otimes e_{n_\cf}^{(\cf)}\otimes \mf\, e_{n_\mf}^{(\mf)}=0.
$$

{ 
From the ladder operators we have constructed we can derive observables useful to quantify the number of healthy and sick cells, and to deduce the general condition of the system.

In particular we first define, see \eqref{24}:
\bea\label{add1}
\Hf=\h^\dag\h,\quad \Cf=\cf^\dagger \cf,\quad \M=\mf^\dagger\mf,
\ena
which act on the elements of $\E$ as follows:
\bea
\nonumber\Hf\varphi_{n_\h,n_\cf,n_\mf}=n_\h \varphi_{n_\h,n_\cf,n_\mf}, \quad \Cf\varphi_{n_\h,n_\cf,n_\mf}=n_\cf \varphi_{n_\h,n_\cf,n_\mf}, \quad \M\varphi_{n_\h,n_\cf,n_\mf}=n_\mf \varphi_{n_\h,n_\cf,n_\mf}.\\
{}
\ena
Following the general scheme proposed in \cite{bagbook}. these operators will be used to measure the number of specific cells of the system and to check is the medical treatment is active or not. For that we introduce, following \cite{sgh} and \cite{bagannal15}, the following expectation values over the normalized state   of the system:
\bea
\langle \Hf\rangle &=&\left<\frac{\Psi}{\lVert \Psi \rVert},\Hf\frac{\Psi}{\lVert \Psi \rVert}\right>=\left\lVert\h \frac{\Psi}{\lVert \Psi \rVert}\right\rVert^2,\label{213}\\
\langle \Cf\rangle&=&\left<\frac{\Psi}{\lVert \Psi \rVert},\Cf\frac{\Psi}{\lVert \Psi \rVert}\right>=\left\lVert\cf \frac{\Psi}{\lVert \Psi \rVert}\right\rVert^2,\label{214}\\
\langle \Mf\rangle&=&\left<\frac{\Psi}{\lVert \Psi \rVert},\M\frac{\Psi}{\lVert \Psi \rVert}\right>=\left\lVert\mf \frac{\Psi}{\lVert \Psi \rVert}\right\rVert^2,\label{215}
\ena
where 
$\langle \cdot,\cdot\rangle $ is the scalar product in $\Hil$ and $\lVert\cdot\rVert^2=\langle \cdot,\cdot\rangle$.  $\langle \Hf\rangle$ and $\langle \Cf\rangle$ count the number of healthy and sick cells in the system, while $\langle \Mf\rangle$ indicates whether a medical treatment is active on $\Sc$, or not. It is easy to show that, for instance, $ \langle \Hf\rangle\leq N_\h-1$:
$$
0\leq \langle \Hf\rangle=\frac{\sum_{n_\h,n_\cf,n_\mf} n_\hf|c_{n_\h,n_\cf, n_\mf}|^2}{\sum_{n_\h,n_\cf,n_\mf}|c_{n_\h,n_\cf, n_\mf}|^2}\leq
\frac{\sum_{n_\h,n_\cf,n_\mf} (N_\hf-1)|c_{n_\h,n_\cf, n_\mf}|^2}{\sum_{n_\h,n_\cf,n_\mf}|c_{n_\h,n_\cf, n_\mf}|^2}=  N_\hf-1.
$$
Hence $\langle \Hf\rangle$  always returns a finite measure of the expected number of healthy cells, which does not exceed the maximum value $N_\hf-1$. Similar considerations lead to the inequalities $\langle \Cf\rangle\leq N_\cf-1$ and $\langle \Mf\rangle\leq N_\mf-1$.

It is useful to introduce the following density-like operators:
\bea
\hat{{P}}_\h=\sum_{n_\h,n_\mf}|\varphi_{n_\h,0,n_\mf}\rangle\langle\varphi_{n_\h,0,n_\mf}|,\quad
\hat{{P}}_\cf=\sum_{n_\cf,n_\mf}|\varphi_{0,n_\cf,n_\mf}\rangle\langle\varphi_{0,n_\cf,n_\mf}|,\label{216}
\ena
where  $|g\rangle\langle h|$, $g,h\in\Hil$, is the rank one operator
which acts on any $f\in \Hil$ as $(|g\rangle\langle h|)f=\langle h,f\rangle \,g$.
The operator  $\hat{{P}_\h}$ projects the state of the system in a subspace of $\Hil$ in which
there are no tumor cells  (the  {\em quantum number} for the sick cells is $n_\cf=0$),  a {\em healthy state}.
For instance, if the state $\Phi_\h$ of the system is a superposition
of only healthy states, $\Phi_\h=\sum_{n_\h,n_\mf}c_{n_\h,n_\mf}\varphi_{n_\h,0,n_\mf}$,
we have $\hat{{P}_\h} \Phi_\h=\Phi_\h$, whereas if we consider a superposition of only {\em sick states},
$\Phi_\cf=\sum\limits_{n_\h,n_\cf>0,n_\mf}c_{n_\h,n_\cf, n_\mf}\varphi_{n_\h,n_\cf,n_\mf}$,
then $\hat{{P}_\h}\Phi_\cf=0$.
Hence we use this operator to obtain a probabilistic measure of the presence of only healthy cells trough the expectation value
\bea
\langle \hat P_\hf\rangle=\left<\frac{\Psi}{\lVert \Psi \rVert},\hat{{P}_\h}\frac{\Psi}{\lVert \Psi \rVert}\right>=
\sum_{n_\h,n_\mf}\left|\left<\varphi_{n_\h,0,n_\mf},\frac{\Psi}{\lVert \Psi \rVert}\right>\right|^2,\label{217}\ena
computed over the normalized state  of the system.
Straightforward computations give
$$
0\leq \langle \hat P_\hf\rangle=\frac{\sum_{n_\h,n_\mf} |c_{n_\h,0, n_\mf}|^2}{\sum_{n_\h,n_\cf,n_\mf}|c_{n_\h,n_\cf, n_\mf}|^2}\leq1,
$$
which motivates why we can assign to $\langle \hat P_\hf\rangle$ the probabilistic interpretation suggested above.
Analogously, $\hat{{P}_\cf}$ projects the state in a {\em sick superposition}, (the { quantum number} of the healthy cells is $n_\h=0$).
The expectation value
\bea
\langle \hat P_\cf\rangle=\left<\frac{\Psi}{\lVert \Psi \rVert},\hat P_\cf\frac{\Psi}{\lVert \Psi \rVert}\right>=
\sum_{n_\cf,n_\mf}\left|\left<\varphi_{0,n_\cf,n_\mf},\frac{\Psi}{\lVert \Psi \rVert}\right>\right|^2=\frac{\sum_{n_\h,n_\mf} |c_{0,n_\cf, n_\mf}|^2}{\sum_{n_\h,n_\cf,n_\mf}|c_{n_\h,n_\cf, n_\mf}|^2},\label{220}\ena
satisfies the inequality $0\leq \langle \hat P_\cf\rangle\leq1$, and for this reason can be considered as the probability to have only tumor cells in $\Sc$.

Extending formulas \eqref{213}-\eqref{215} above for the positive hermitian operators $\Hf$, $\Cf$ and $\Mf$ to a generic, not necessarily hermitian operator $\hat{\mathcal{O}}$, we introduce here
\bea
\left|\langle \hat{\mathcal{O}}\rangle\right| =\left|\left<\frac{\Psi}{\lVert \Psi \rVert},\hat{\mathcal{O}}\frac{\Psi}{\lVert \Psi \rVert}\right>\right|,\label{221}
\ena
which will be used in the following to get some interesting information on the system\footnote{Of course, if $\hat{\mathcal{O}}$ coincides, say, with $\Hf$, \eqref{221} coincides with \eqref{213}.}.

\section{The Hamiltonian of the system}\label{sec::hamiltonian}

We have introduced before the main ingredients of our system $\Sc$: the healthy ($\h$) and the sick ($\cf$) cells, and the medical treatment $(\mf)$. Notice that, compared with \cite{haven2}, we are not inserting here any factor which causes the transition of a cell from a healthy to a sick state, while we are considering explicitly some medical treatment which should contrast the sickness. The reason for not inserting any carcinogenic factor is that we are already assuming that this transition occurs, and this  implies the presence of such a factor, which needs not to be considered as a dynamical variable of $\Sc$.  This has nice consequences on the dimensionality of our Hilbert space, which is somehow lowered by this choice, improving in this way the computational time. We are much more interested in discussing the changes in the reaction of the tumor cells depending on when and how the medical treatment we consider acts on $\Sc$. We will discuss this aspect in many details in Section \ref{sec::numerics}.

 The mechanisms which we imagine in $\Sc$ are the following: first of all,    healthy cells become sick ($\h\rightarrow \cf$). Secondly, these sick cells multiply by mitosis ($\cf\rightarrow \cf+\cf$). Healthy cells multiply too, but possibly with a lower frequency ($\h\rightarrow \h+\h$). We also imagine two different kind of medical treatments: in the first one, the medicine disappears during the treatment which destroys a sick cell ($\cf+\mf\rightarrow \emptyset$). In the second treatment the medicine acts but it does not disappear: $\cf+\mf\rightarrow \mf$. The first is when, for instance, the medical treatment is active only for some time, and then is removed, while the second is more connected to some continuous (in time) treatment. {We will see that the second treatment is much more efficient than the first one, which turns out to be not particularly useful, in the long run.} 

As usual, all these mechanisms will be encoded by a suitable Hamiltonian $H$. However, modifying the standard approach, \cite{bagbook}, the Hamiltonian which we propose is manifestly not hermitian. And, as we have discussed in the Introduction, the lack of hermiticity will not be caused by the presence of some complex-valued parameter in $H$, but by the absence of some {\em hermitian conjugate} terms in the Hamiltonian itself. To be more explicit, we first introduce the following operator:
\be\label{htilde}
\left\{
\begin{array}{ll}
	\tilde H=\tilde H_0+\tilde H_I+\tilde H_g\\
	\tilde H_0=\omega_{\h} \Hf+\omega_\cf \Cf+\omega_\mf \M,\\
	\tilde H_I=\mu_{\h \cf}(\h \Hf\cf^\dag+\cf \Hf\h^\dagger)+\mu_{\h \h}(\h^\dag\Hf+\Hf\h)+\mu_{\cf\cf}(\cf^\dag\Cf+\Cf\cf),\\
	\tilde H_m=\mu_{\cf\mf_1}(\cf\Cf\mf+\mf^\dagger \Cf\cf^\dagger)+\mu_{\cf\mf_2}(\cf\Cf+\Cf\cf^\dagger) \M,
\end{array}
\right.
\en
where the parameters $\omega_\alpha$ and $\mu_{\alpha\beta}$ are assumed to be real and, for the moment,  time independent. Hence $\tilde H=\tilde H^\dagger$. The meaning of $\tilde H_0$ is well understood, \cite{bagbook}: first of all, it  creates no interesting dynamics when all the $\mu$'s are zero. In fact, in this case, $\Hf$, $\Cf$ and $\M$ commute with $\tilde H$, and therefore they all stay constant in time. This means that, for instance, if we start with a situation where there are no cancer cells, then their number remains zero for all $t\geq0$. However, when some of the $\mu$'s are different from zero, the values of the $\omega_\alpha$ describes a sort of inertia of the agent $\alpha$: the larger this value, the smaller the variations of the mean value of the related number operator. This effect was observed in several applications, see \cite{bagbook}.

{
The first part of the first term in $\tilde H_I$, $\h\Hf\cf^\dag$, describes the mutation of healthy  into sick cells. This is because one healthy cell is destroyed by $\h$ and a sick cell is created by $\cf^\dagger$. Of course this mutation is more probable when $\Sc$ is made of many healthy cells, and this explains the appearance of the number operator $\Cf$, which works by counting the number of healthy cells. Its hermitian conjugate, $\cf\Hf\h^\dagger$, describes the opposite transition, from a cancer  to a healthy cell. It is clear that this effect is like a spontaneous healing, which is quite unexpected and  not well recognized from a  biological point of view.

Something similar can be repeated for the other terms in $\tilde H_I$ and for those in $\tilde H_m$: the terms  $\h^\dag \Hf$ and  $\cf^\dag \Cf$  describe duplications of the healthy and of the sick cells, respectively. This duplication can only occur, of course, if some cell of that particular kind already exists, and is more frequent if these cells are many. This is modelled by the presence of  $\Hf$ and $\Cf$ in the Hamiltonian, respectively.
Of course, since we expect that the duplication is faster for the sick rather than for the healthy cells, it is natural to take $\mu_{\h\h}<\mu_{\cf\cf}$.
Once again,  their hermitian conjugates 
$ \Hf\h$ and  $ \Cf\cf$ are {\em biologically strange}. In fact, they would correspond to the death of the healthy and of the cancer cells. This could be understood as the effect of age, for instance, or, again, of some spontaneous healing, but in any case it is quite strange to imagine that the, say, cancer cells are created and destroyed in $\Sc$ at the same rate.

Similar difficulties arise also when trying to interpret $\tilde H_m$.
The term $\cf\Cf\mf$ describes the death of  sick cells as a consequence of some medicinal treatment, which disappears after being used.
Similarly,  $\cf\Cf\Mf$ describes the death of  sick cells, but the treatment stays active since $\mf$ is replaced by $\Mf$. As stated above, these two terms reflects two different possible treatments of the disease. But $H_m$ also contains two terms which are not quite natural, but are needed if we want to work with hermitian Hamiltonians: $\mf^\dagger\Cf \cf^\dagger$ describes a situation in which, in presence of medicine, both the number of cancer cells and the amount of medicine increase. It is like if the system evolves backward in time. And the same could be deduced from  $\cf^\dagger \Cf \Mf$: the number of sick cells increases, even if the medical treatment is active.}

We conclude that, also if some of the contributions above can be somehow motivated,  this motivation is not always really satisfying
and induces biological effects not easy to understand or not observed in real tumoral growth.
For this reason, in this paper, we are considering the possibility to work with a non-hermitian Hamiltonian, extracting from $\tilde H$ only those terms which are meaningful, and  reasonable, from a purely biological point of view. This is, in fact, a big innovation with respect to what was done so far, in similar contexts.

This suggests to define the following Hamiltonian:
\be\label{31}
\left\{
\begin{array}{ll}
	 H= H_0+ H_I+ H_g\\
	 H_0=\omega_\h \Hf+\omega_\cf \Cf+\omega_\mf \Mf,\\
	 H_I=\mu_{\h\cf}\h\Hf\left(\cf^\dag+\hat P_{N_\cf}\right)+\mu_{\h\h}\h^\dag\Hf+\mu_{\cf\cf}\cf^\dag\, \Cf,\\
	 H_m=\mu_{\cf\mf}\cf\Cf\,\Mf.
\end{array}
\right.
\en

\vspace{2mm}

{\bf Remark:--} With respect to $\tilde H_m$, in $H_m$ we are not considering $\mu_{\cf\mf_1}\cf\Cf\mf$, that is we are neglecting the possibility of using medicines which disappear during the treatment. The 
reason is that our numerical investigations gave evidence that this term does not really modify the dynamics of the system. This can be mathematically understood because of the presence of $\mf$ in the term above: the (repeated) action of $\mf$ over a generic vector of $\Sc$ destroys the state! From the point of view of its biological interpretation, it is like if the medicine stays active just for a while, and then it is consumed. It is clear that this approach cannot be particularly efficient, compared with a different (continuous) treatment. This is exactly what our simulations show.  For this reason, we only consider only the second treatment, and, to simplify the notation, we put $\mu_{\cf\mf}=\mu_{\cf\mf_2}$.

\vspace{2mm}

Notice that $H_I$ contains a correction to the original mutation term $\h\Hf\cf^\dag$ in \eqref{htilde}, based on the following operator:

\bea
\hat P_{N_\cf}:=\sqrt{N_\cf}\sum\limits_{n_\hf,n_\mf}|\varphi_{n_\hf,N_\cf-1,n_\mf}\rangle\langle\varphi_{n_\hf,N_\cf-1,n_\mf}|.
\label{33}\ena
The rationale for adding this term is the following: $P_{N_\cf}$ has a non zero effect only on those states whose expansion in terms of $\E$ contains some vector $\varphi_{n_\hf,N_\cf-1,n_\mf}$. Otherwise its action is trivial. In particular we see that, while $\cf^\dagger\varphi_{n_\hf,N_\cf-1,n_\mf}=0$, $(\cf^\dagger+P_{N_\cf})\varphi_{n_\hf,N_\cf-1,n_\mf}=\varphi_{n_\hf,N_\cf-1,n_\mf}$. This is useful  to describe a sort of equilibrium when the maximum number of sick cells is reached, avoiding to destroy the system completely during the time evolution, in this particular case: stated differently, if the cells of $\Sc$ are extremely sick, then they stay sick!

The Hamiltonian \eqref{31} is manifestly non hermitian. This implies that we have to impose some choice on how to derive the time evolution of the system. The reason for this is the following: if $H\neq H^\dagger$, the Schr\"odinger and the Heisenberg equations are not equivalent any more. This is widely discussed, for instance, in \cite{bagbook}, and is mainly based on the fact that the operator $e^{-iHt}$ is not unitary. The choice we adopt here is quite common, for instance in quantum optics: we assume that, even if $H\neq H^\dagger$, the time evolution of the wave function of $\Sc$ is still driven by the Schr\"odinger equation of motion
\be
i\dot{\Psi}(t)=H\Psi(t).
\label{32}
\en
Its solution is $\Psi(t)=e^{-iHt}\Psi(0)$, where $\Psi(0)$ is the initial state of the system. In our case, $\Psi(0)$ describes how many healthy and sick cells exist at $t=0$, and if the medical treatment is also active in the system or not, when the evolution starts. In general, $\Psi(0)$ can be expressed in terms of the vectors in $\E$ as in
\eqref{210}.

We assume from now that $N_\mf=1$, which essentially means that the medical treatment is  a dichotomous variable: the   eigenvalues of $\Mf$ can only be zero and one: absence or presence of a medical treatment.
{In our computations we will always take $N_\cf>N_\h$, since the biology of cancer cells suggests that the sick cells are luckily to grow, in number, more than the healthy ones.} 
 More explicitly, we fix $N_h=50$ and $N_s=150$, 
much less than in real experimental settings.
However, these  values are a good compromise between a realistic situation and a reasonable computational time.

\subsection{Evolution of the system}
We are now ready to derive the differential equations ruling the time evolution of $\Sc$.
To simplify the notation we introduce, when possible, the vector $\n=(n_\h,n_\cf,n_\mf),\quad n_\h=0,\ldots,N_\h-1,\quad n_\cf=0,\ldots,N_\cf-1, \quad n_\mf=0, 1= N_\mf-1$.
To determine the densities , we need first to compute  the time evolution of the state of the system.
In particular, let
$\Psi(0)=\sum\limits_{n_\h,n_\cf,n_\mf} c_{\n}(0) \varphi_{\n}$ be the initial state of the system,
where the complex scalar coefficients $c_{\n}(0)$ can be   chosen to satisfy
$\sum\limits_{n_\h,n_\cf,n_\mf}|c_{\n}(0)|^2=1$ (but not necessarily). In what follows we will always assume that  $c_{\n}(0)=\delta_ {\n,\n^o}$, where $\n^o=(n_\h^o,n_\cf^o,n_\mf^o)$. This means that, at $t=0$, $\Sc$ is in a common eigenstate of the operators in (\ref{add1})
We have already stressed that, due to the non hermiticity of our Hamiltonian \eqref{31}, the evolution is non unitary, so that,
in general, $\|\Psi(t)\|\neq\|\Psi(0)\|$ for $t>0$

The time evolution of $\Psi(0)$  is driven by the
Schr\"odinger equation \eqref{32}, where the time dependence is all contained in the coefficients
$c_{\n}$, which therefore are functions of time, $c_{\n}(t)$. Using the orthogonality conditions of the basis vectors $\varphi_{\n}$, we obtain the following system
\be
i\frac{d c_{\n}(t)}{d t}=\langle\varphi_{\n},H\Psi(t)\rangle,\quad \forall \n,
\label{SHnl22Dext_a}
\end{equation}
which produces the following set of differential equations, $\forall n_\hf,n_\cf,n_\mf$: 
%
%
\beano
&&i\frac{d c_{n_\hf,n_\cf,n_\mf}(t)}{d t}=\Big(\omega_\hf n_\hf+\omega_\cf n_\cf+\omega_\mf n_\mf\Big) c_{n_\hf,n_\cf,n_\mf}+\\
&&\Big([n_\hf<N_{\hf}-1][n_\cf>0]\mu_{\hf\cf}\left(n_{\hf}+1\right)\sqrt{(n_{\hf}+1)n_{\cf}}c_{n_\hf+1,n_\cf-1,n_\mf}+\\
&&[n_\hf<N_{\hf}-1][n_\cf=N_\cf-1]\mu_{\hf\cf}(n_{\hf}+1)\sqrt{(n_{\hf}+1)N_{\cf}}c_{n_\hf+1,N_\cf-1,n_\mf}+\\
&&\Big[n_\hf>0\Big]\mu_{\hf\hf}(n_{\hf}-1)\sqrt{n_{\hf}}c_{n_\hf-1,n_\cf,n_\mf}+\Big[n_\cf>0\Big]\mu_{\cf\cf}(n_{\cf}-1)\sqrt{n_{\cf}}c_{n_\hf,n_\cf-1,n_\mf}\Big)+\\
&&\Big(\Big[n_\cf<N_{\cf}-1\Big]\mu_{\cf\mf}n_{\mf}(n_{\cf}+1)\sqrt{n_{\cf}+1}c_{n_\hf,n_\cf+1,n_\mf}\Big),
\enano
where $\Big[\bullet\Big]$ is a logical operator returning $1$ if $\bullet$ is true, and $0$ otherwise.

This is a linear system of ordinary differential equations if all the parameters of $H$ are fixed. In this case, a solution can be easily deduced, in principle. However, what is more interesting for us is to consider some of these parameters dependent on the density of the (sick and/or healthy) cells. This is relevant for us, since it describes the fact that the medication starts when the tumour has grown to a certain size, and not before, possibly because its presence has not even be recognized. We are particularly interested in analysing what happens when we modify this parameter, changing the instant in which the medication starts, and its strength. In this case  the above system becomes, in general, nonlinear, since the parameters   depend on the mean values of the number operators in (\ref{213})-(\ref{215}), and its solution is not so simple and needs to be computed numerically. In particular, the numerical solutions we shall adopt in this work are obtained by using an explicit Runge-Kutta  formula based on the Dormand-Prince pair, \cite{DP80},
where  densities in \eqref{213}-\eqref{215} are computed at each time step. 

%
%

\section{Numerical results}
\label{sec::numerics}
In this Section we present the numerical outcomes of our model. We 
consider two main scenarios: absence and presence of medical treatment. And,  in the latter scenario, we propose different strategies.

In many simulations  we  fix the parameter $\mu_{\h\cf}=1$ in \eqref{31}, which is equivalent to fix the time scale according to  $\mu_{\h\cf}$. We always consider a time dependent production of the healthy cells, assuming that this production degrades more and more in presence of higher number of tumor cells. This  reflects real phases of tumor cells proliferation, \cite{Tum1}, which removes vital space to the healthy cells.
In particular we introduce a specific dependence of $\mu_{\hf\hf}$ on $\langle \hat N_{\cf}\rangle$, by choosing  a logistic--like expression $\mu_{\hf\hf}(\langle \hat N_{\cf}\rangle):=\tilde\mu_{\hf\hf}\left(1-\frac{\langle \hat N_{\cf}\rangle}{{N}_{\cf}-1}\right)$. Here
$\tilde\mu_{\hf\hf}$ is the maximum value assumed by $\mu_{\hf\hf}$ when no tumor cell is present, $\langle \hat N_{\cf}\rangle=0$, and $\mu_{\hf\hf}(\langle \hat N_{\cf}\rangle)$ decreases for increasing $\langle \hat N_{\cf}\rangle$.
Finally, recalling that the parameters $\omega_{\hf,\cf,\mf}$  measure the resistance of the system to change, \cite{bagbook}, we make this resistance low
by taking very small values of them, $\omega_{\hf,\cf,\mf}=10^{-2}$.

\subsection{No medical treatment}

We start considering the scenario in which no medical treatment is active on $\Sc$.
This implies that $\mu_{\cf\mf}=0$. 
We have considered first three different configurations, in which  $\mu_{\h\cf}=1$, always:
$\mu_{\cf\cf}=0.5,\tilde\mu_{\hf\hf}=0.25$ in the  configuration $R1$, $\mu_{\cf\cf}=2,\tilde\mu_{\hf\hf}=1$ in configuration $R2$, and $\mu_{\cf\cf}=0.125,\tilde\mu_{\hf\hf}=0.0625$ in  configuration $R3$.
Scenarios $R1$ and $R2$ represent respectively situations in which the relevant effects are mutation of healthy into sick cells  ($R1$), and  proliferation of tumor cells  ($R2$). In $R3$,  proliferation of both tumor and healthy cells is a weak effect when compared to the mutation. 

Initial conditions for all the simulations are $\langle \hat N_{\hf}\rangle=N_\hf-1,\langle \hat N_{\cf}\rangle=0$, corresponding to 
$\Psi(0)=\varphi_{N_\hf-1,0,0}$. 
This  initial state consists in only healthy cells  in the system. Of course, the condition $\mu_{\cf\mf}=0$ implies that no medical treatment is active on $\Sc$.

The outcomes in terms of number  of healthy, $\langle \hat N_{\hf}\rangle$, and tumor cells, $\langle \hat N_{\cf}\rangle$, are shown in Figure \ref{NoMed_N}, while the probability of having only healthy, $\langle \hat P_{\hf}\rangle$, or only tumor cells, $\langle \hat P_{\cf}\rangle$, are shown in Figure \ref{NoMed_P}.
We observe different stages in  the three scenarios.
In a first stage, as expected, the number of healthy cells decreases, as a consequence of their mutation into tumor cells. 
This leads to a very fast decrease of the probability $\langle \hat P_\hf\rangle$ of having only healthy cells, which rapidly decreases to zero in a very small time interval $\approx 0.01$, Figure \ref{NoMed_P}.
A second stage is mainly ruled by the proper proliferation of tumor cells by mitosis, with the further decay of the number of healthy cells.
We can observe the beginning of this second stage in  Figure \ref{NoMed_N}: at  some specific time, depending on the configuration considered, the number of tumor cells
suddenly  growths while the number of healthy cells decreases.
 This is clearly visible at $t\approx 0.046, 0.11$ for the configurations $R1$ and $R2$.
In the final stage the tumor cells saturate and essentially cover the whole system, with an increasing probability of having only tumor cells, see Figure \ref{NoMed_P},  while the number of healthy cells tends asymptotically to zero.

The various stages described above can be conveniently understood by means of the analysis of the expectation values of the various terms in the Hamiltonian \eqref{31}.
In particular, we compute as in \eqref{221} the moduli $|\langle H_{\hf\cf}\rangle|, |\langle H_{\hf\Hf}\rangle|,|\langle H_{\cf\Cf}\rangle|$ of the expectation values  related to the mutation term $\mu_{\h\cf}\hf\Hf\cf^\dag$, and to the proliferation terms $\mu_{\hf\hf}\h^\dag\Hf$ and $\mu_{\cf\cf}\cf^\dag\Cf$.
Adapting to the present context the typical meaning associated to the expectation values in quantum mechanics, the higher  these values, the more relevant are the contributions of the related operators to the dynamics of the system.
The time evolutions of  $|\langle H_{\hf\cf}\rangle|, |\langle H_{\hf\Hf}\rangle|,|\langle H_{\cf\Cf}\rangle|$ are shown in Figures \ref{NoMed_H_P1}-\ref{NoMed_H_P2}
for the configuration $R1, R2$, respectively. The first stage is mainly governed by the dynamics induced by  $\mu_{\h\cf}\hf\Hf\cf^\dag$,
whereas the second stage is governed by the tumor cell mitosis induced by $\mu_{\cf\cf}\cf^\dag\Cf$: this effect is particularly evident in the configuration $R2$ where the value of the parameter $\mu_{\cf\cf}$ is higher than in $R1$. At the same time, we can observe the small effect induced by the healthy cell mitosis contribution
$\mu_{\h\h}\h^\dag\Hf$.
{ The last small increment  in $\langle H_{\hf\cf}\rangle$ at $t\approx1.3$ and $t\approx0.6$} for the two configurations, is  due to the increasing  contribution of $\hat P_{N_\cf}$ defined in \eqref{33}.
This contribution becomes relevant when  $\langle \hat N_{\cf}\rangle\approx{N}_{\cf}-1$, which is the case when the state $\Psi(t)$ is essentially  a combination of the states $\varphi_{n_\hf,N_{\cf}-1,n_\mf}$.
At the same time $|\langle H_{\cf\Cf}\rangle|\rightarrow0$, consistently with the fact the no other tumor cells are produced.

\begin{center} 
\begin{figure}            
\subfigure[]{\includegraphics[width=8cm]{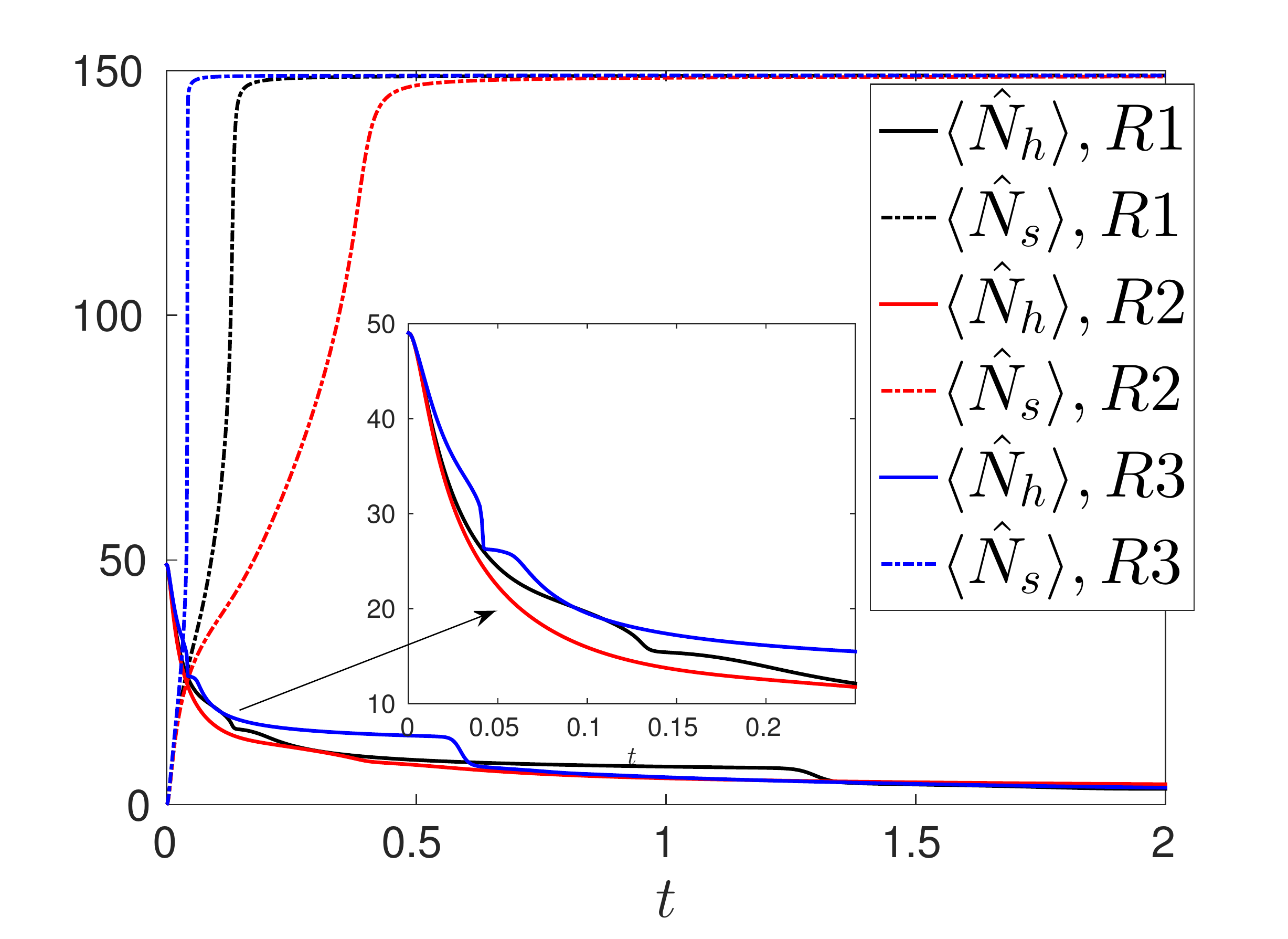}\label{NoMed_N}}
\subfigure[]{\includegraphics[width=8cm]{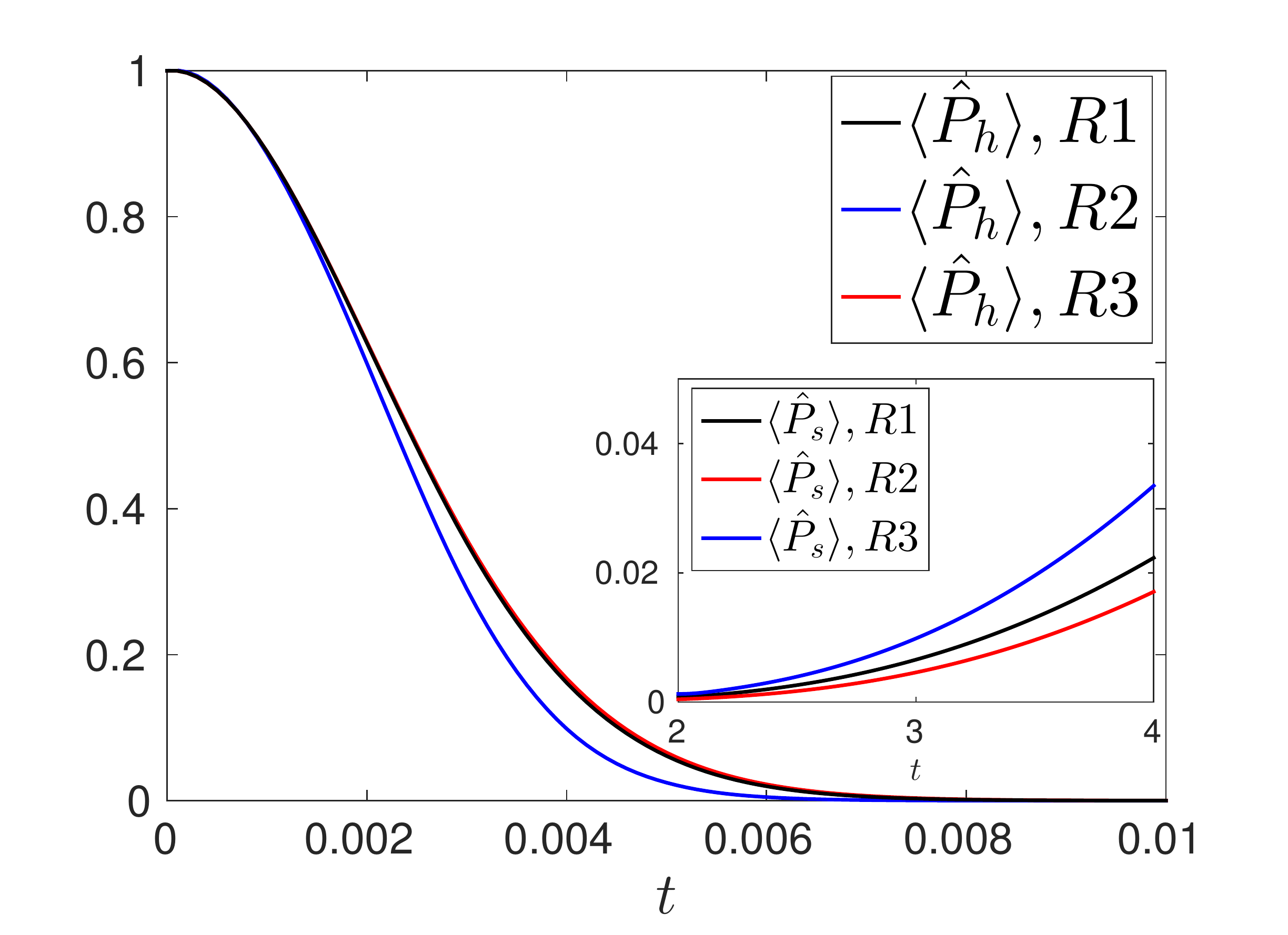}\label{NoMed_P}}
\caption{\textbf{(a)}  Number $\langle \hat N_\hf\rangle,\langle \hat N_\cf\rangle$ of healthy  and tumor cells 
  configurations $R1$, $R2$ and $R3$.
In the inset the magnification of the early time time evolutions of $\langle \hat N_\hf\rangle$.
\textbf{(b)} Probability $\langle \hat P_\hf\rangle$ of having only healthy cells for the same scenarios.
In the inset  the  large time evolutions of the probability $\langle \hat P_\cf\rangle$ of having only tumor cells  .}
\end{figure}
\end{center}

\begin{center} 
\begin{figure}               
\subfigure[Configuration R1]{\includegraphics[width=8cm]{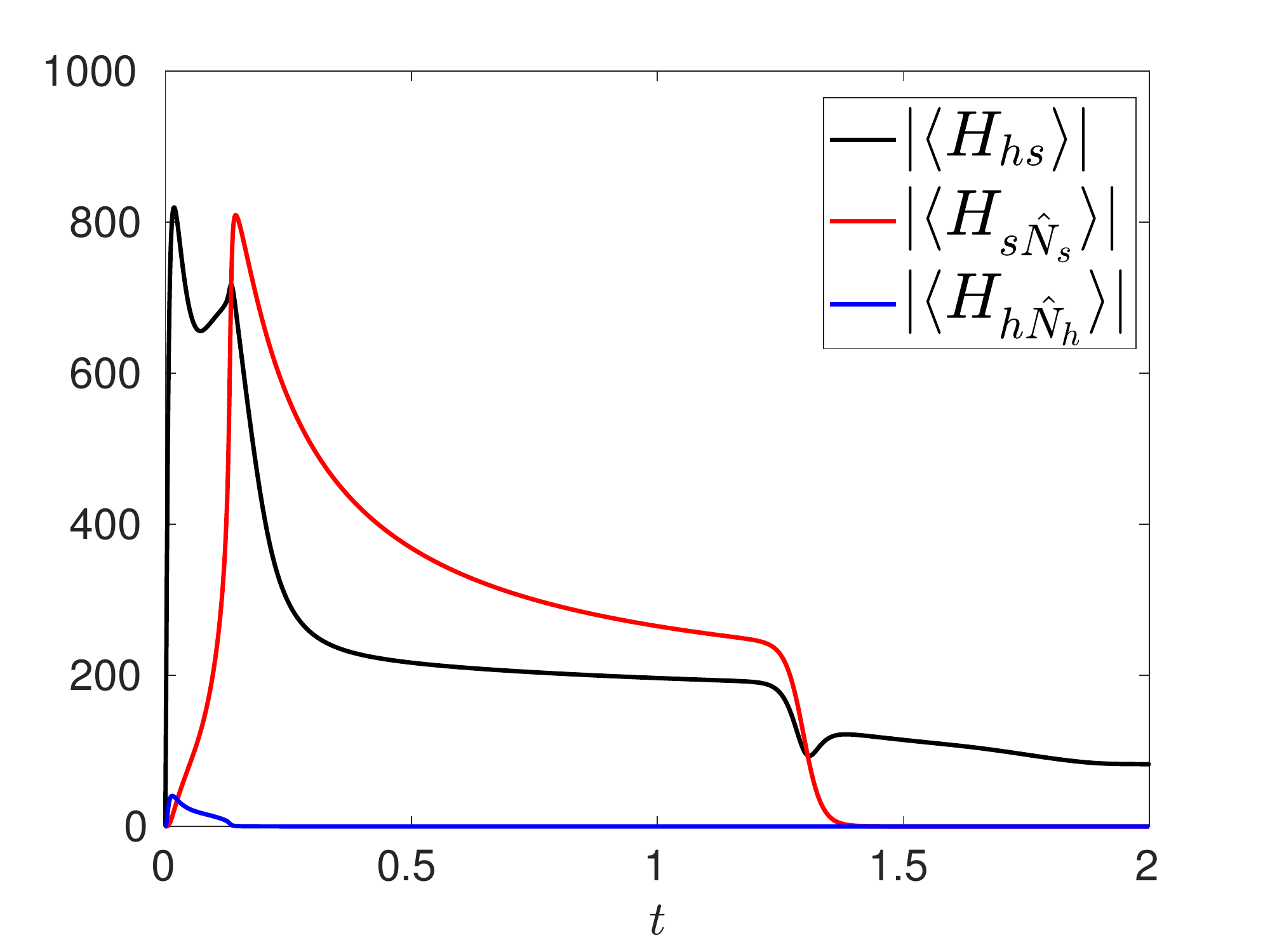}\label{NoMed_H_P1}}
\subfigure[Configuration R2]{\includegraphics[width=8cm]{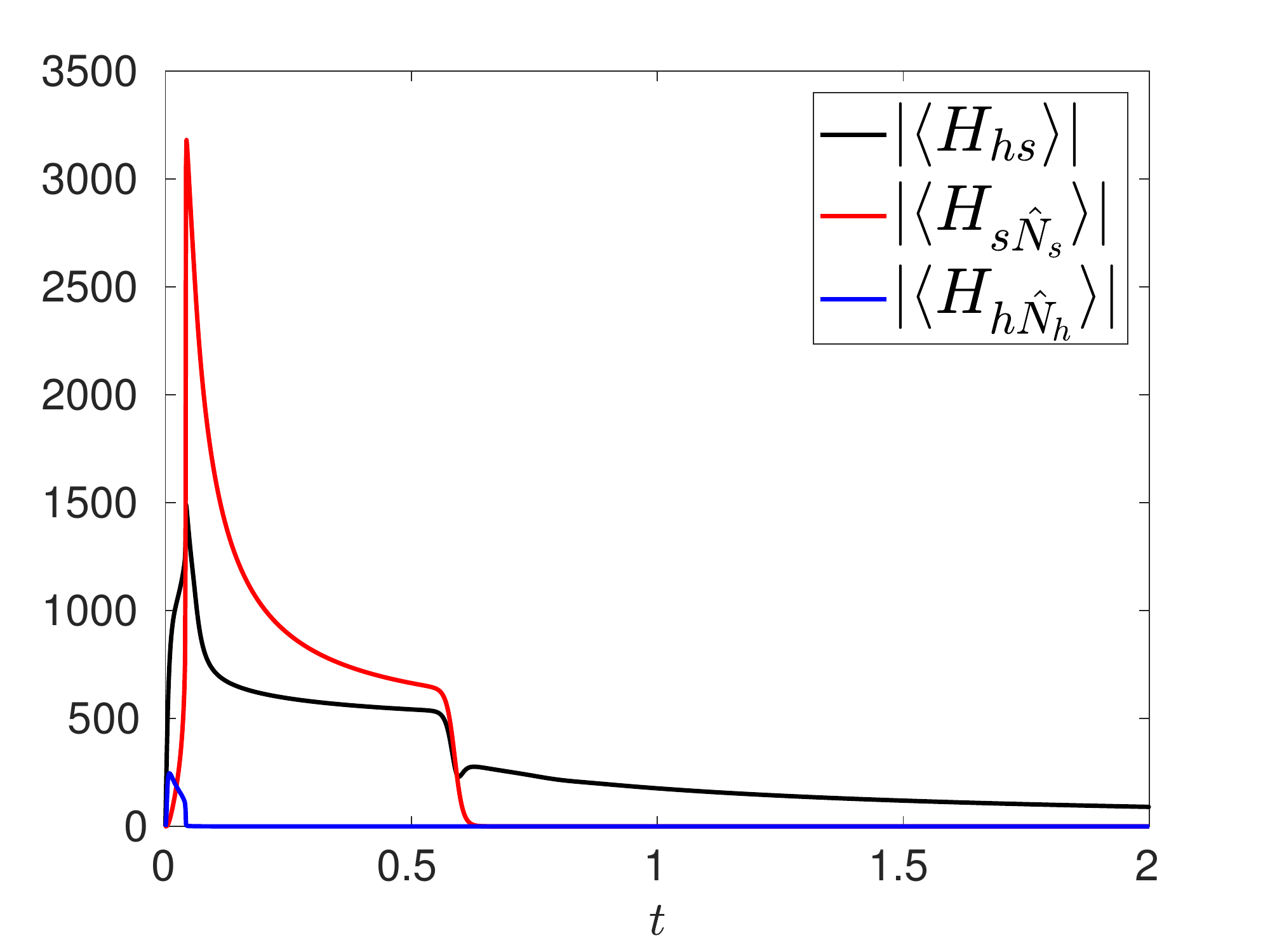}\label{NoMed_H_P2}}
\caption{  Time evolutions of  the moduli  $|\langle H_{\hf\cf}\rangle|, |\langle H_{\cf\Cf}\rangle|,|\langle H_{\hf\Hf}\rangle|$ for the configuration $R1$ \textbf{(a)}, and 
 for the configuration $R2$ \textbf{(b)}.}
\end{figure}
\end{center}

\subsection{Medical treatment I: time independent treatment }

The three scenarios introduced before are mainly meant to show that our model can efficiently describe a {\em global} mutation of healthy into sick cells. This mutation takes some time, of course. The relevant aspect of the model, for us, is whether this process can be stopped or, better, reversed, and how. This is why now we discuss what happens if $\mu_{\cf\mf}$ is taken different from zero. 

First we consider the case in which the medical treatment starts from the beginning of the evolution, with the further assumption that  the parameter $\mu_{\cf\mf}$ is time independent.
Initial conditions and other parameters are the same used in configuration $R1$ considered before.
We show in Figures \ref{Med_sm2_comparison_NH}-\ref{Med_sm2_comparison_NS}  the time evolutions of the mean values $\langle \hat N_{\hf}\rangle,
\langle \hat N_{\cf}\rangle$ of healthy and tumor cells 
for different
values of $\mu_{\cf\mf}$.
The results show that the action of a medical treatment reduces $\langle \hat N_{\cf}\rangle$ 
as the intensity of the treatment is increased, see Figure \ref{Med_sm2_comparison_NS}. This is exactly the result one should expect for biological reasons.
Conversely, the number $\langle \hat N_{\hf}\rangle$  is no more asymptotically going to zero, as in absence of medical treatment, see Figure \ref{NoMed_N}},
 and it   
stabilizes (or oscillates) around a value which increases with $\mu_{\cf\mf}$. This is related to the fact that the tumor cells do not saturate the system, so that the production of
healthy cells, given by $\mu_{\hf\hf}\h^\dag\Hf$, does no more vanish. $\Sc$ reaches a sort of equilibrium between healthy and sick cells: neither the first, nor the second, completely disappear from the system.

The intensity of the medical treatment during time is  related to measure of the intensity  of $\mu_{\cf\mf}\cf\Cf\Mf$, $|\langle H_{\cf\mf}\rangle|$, shown in
Figure \ref{Med_Hsm_comparison}. We observe that  $|\langle H_{\cf\mf}\rangle|$ does not  increase with $\mu_{\cf\mf}$ at all time: several oscillations are observed, arising from the various relations between the agents of the system. Notice that, for all values of  $\mu_{\cf\mf}$, $|\langle H_{\cf\mf}\rangle|$ is much larger than $|\langle H_{\hf\Hf}\rangle|$, see Figures \ref{Med_Hhh_comparison} and \ref{Med_Hsm_comparison}.

%
%

We observe that the probability $\langle \hat P_\cf \rangle$ of having only tumor cells in the system is always vanishing, reaching at most a value of order $10^{-7}$, well below what we have found in absence of medical treatment, see Figure \ref{NoMed_P}.

\begin{center} 
\begin{figure}
\subfigure[]{\includegraphics[width=8cm]{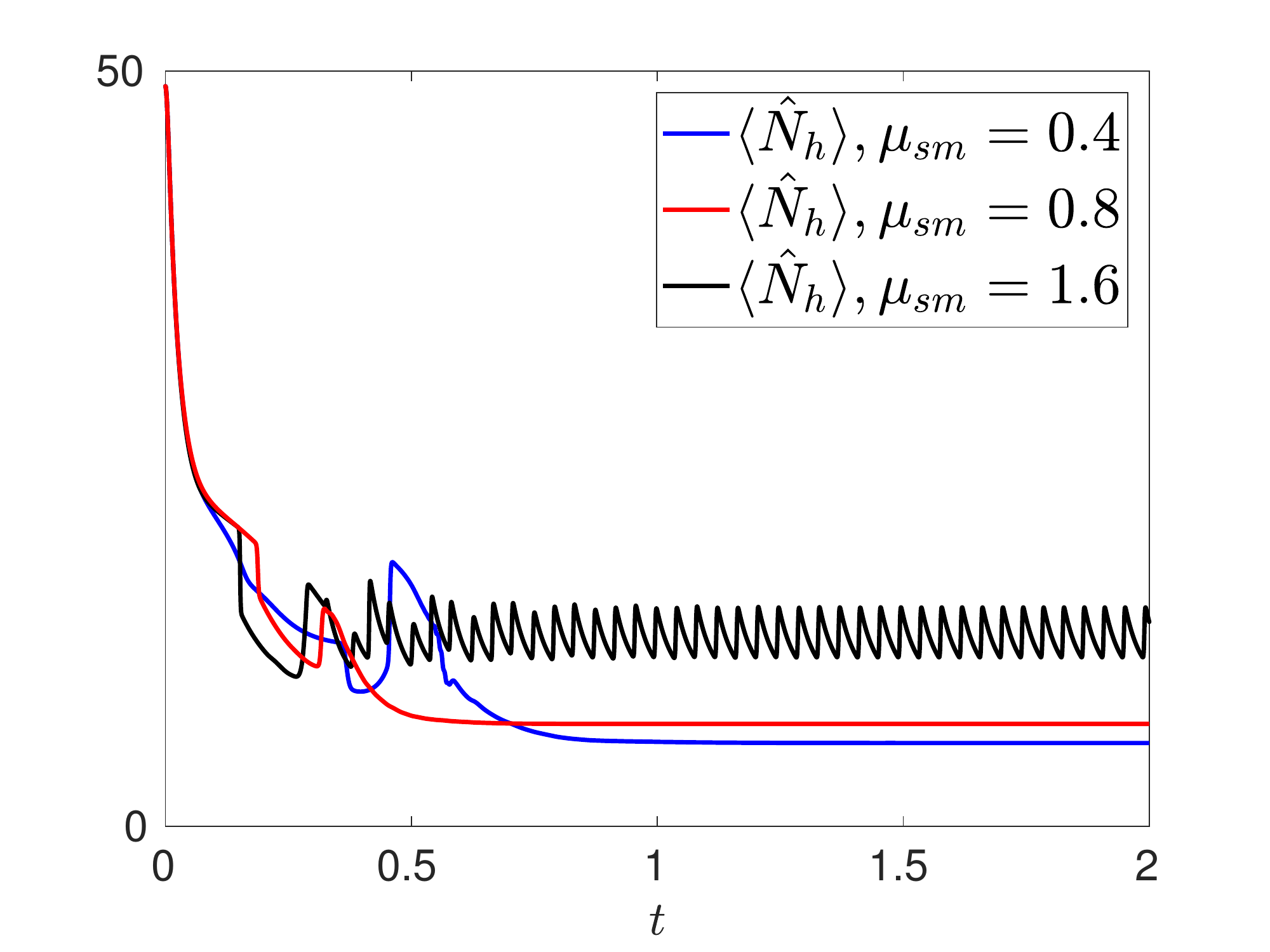}\label{Med_sm2_comparison_NH}}
\subfigure[]{\includegraphics[width=8cm]{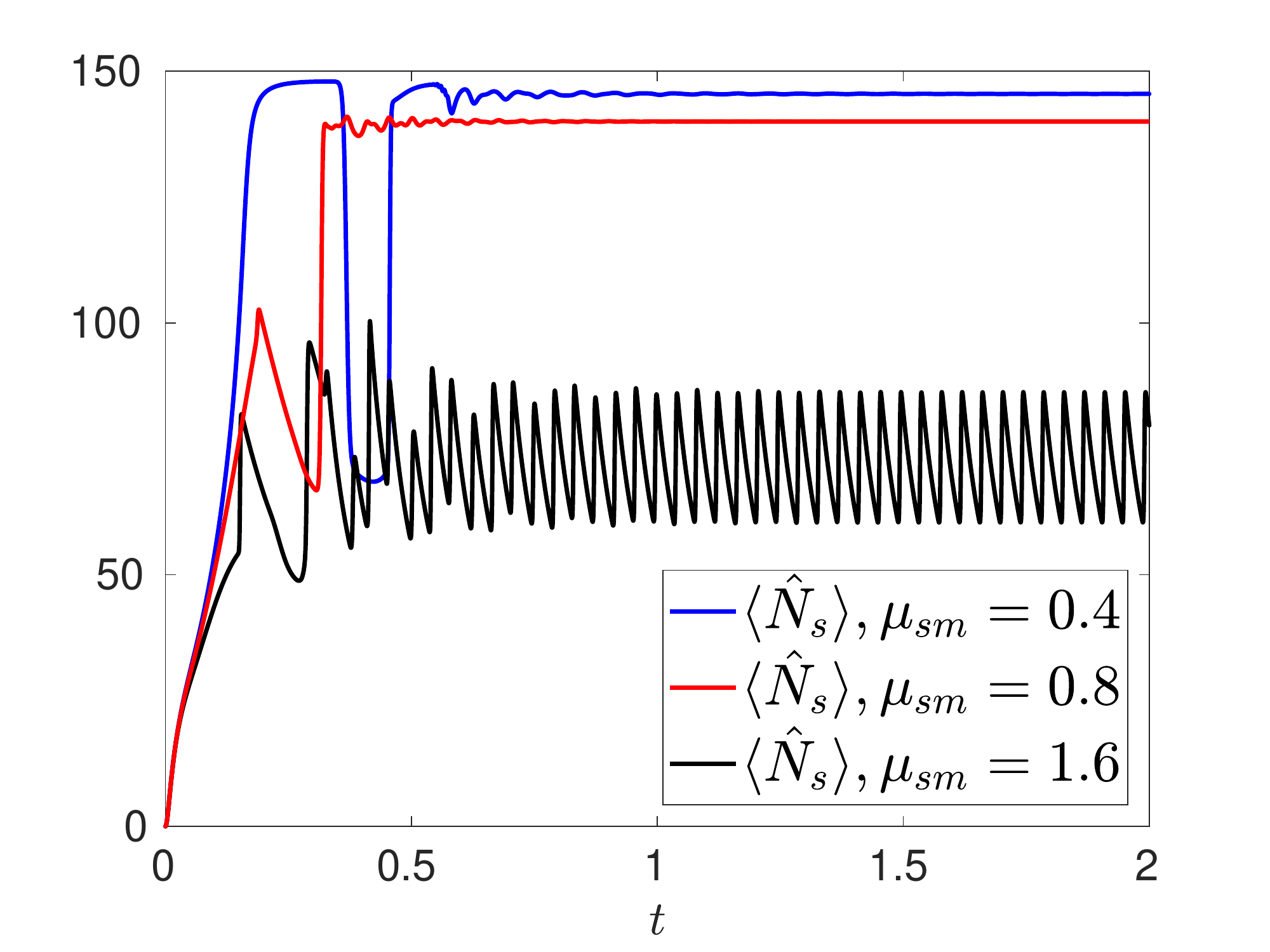}\label{Med_sm2_comparison_NS}}     
\subfigure[]{\includegraphics[width=8cm]{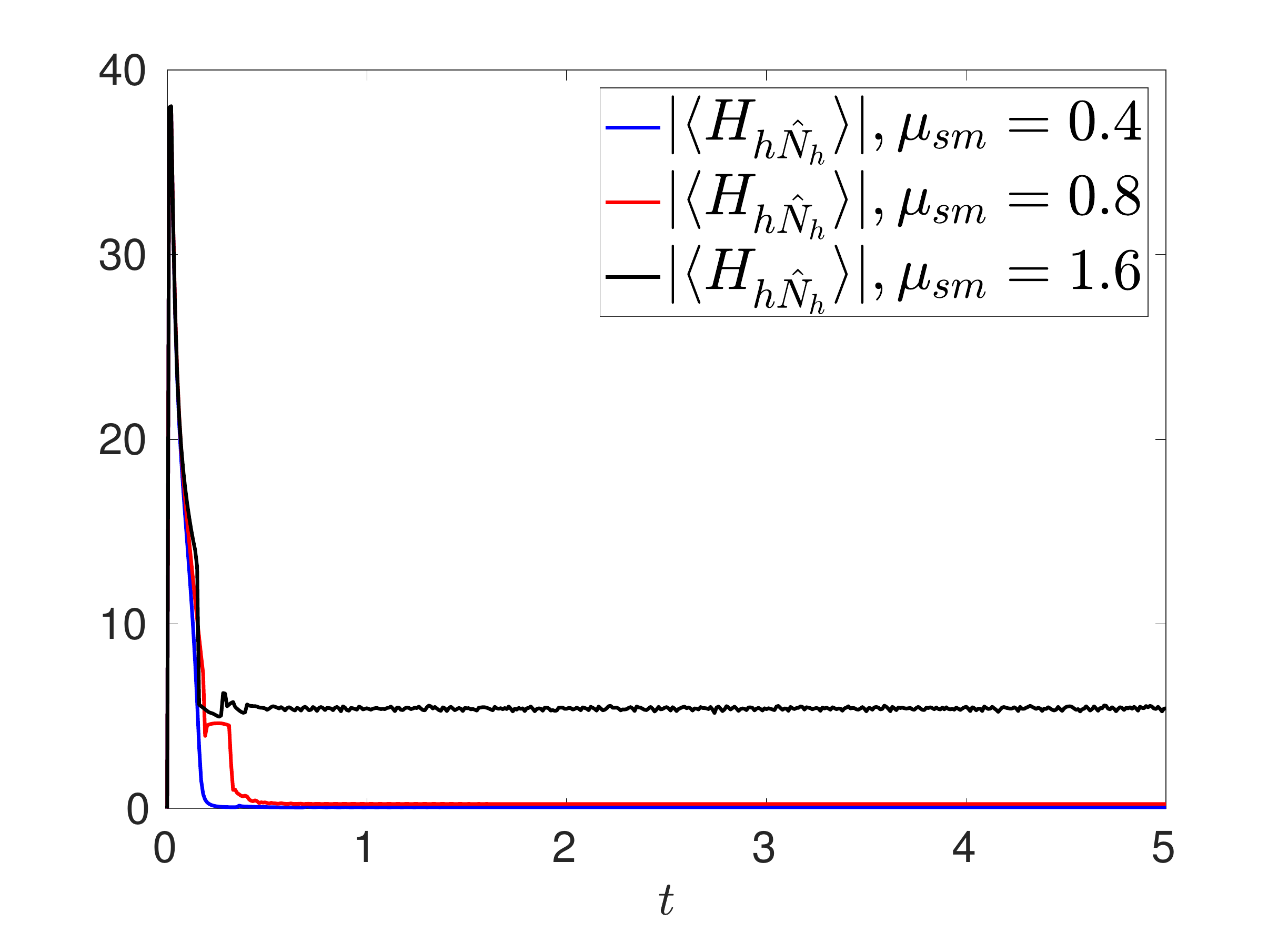}\label{Med_Hhh_comparison}}
\subfigure[]{\includegraphics[width=8cm]{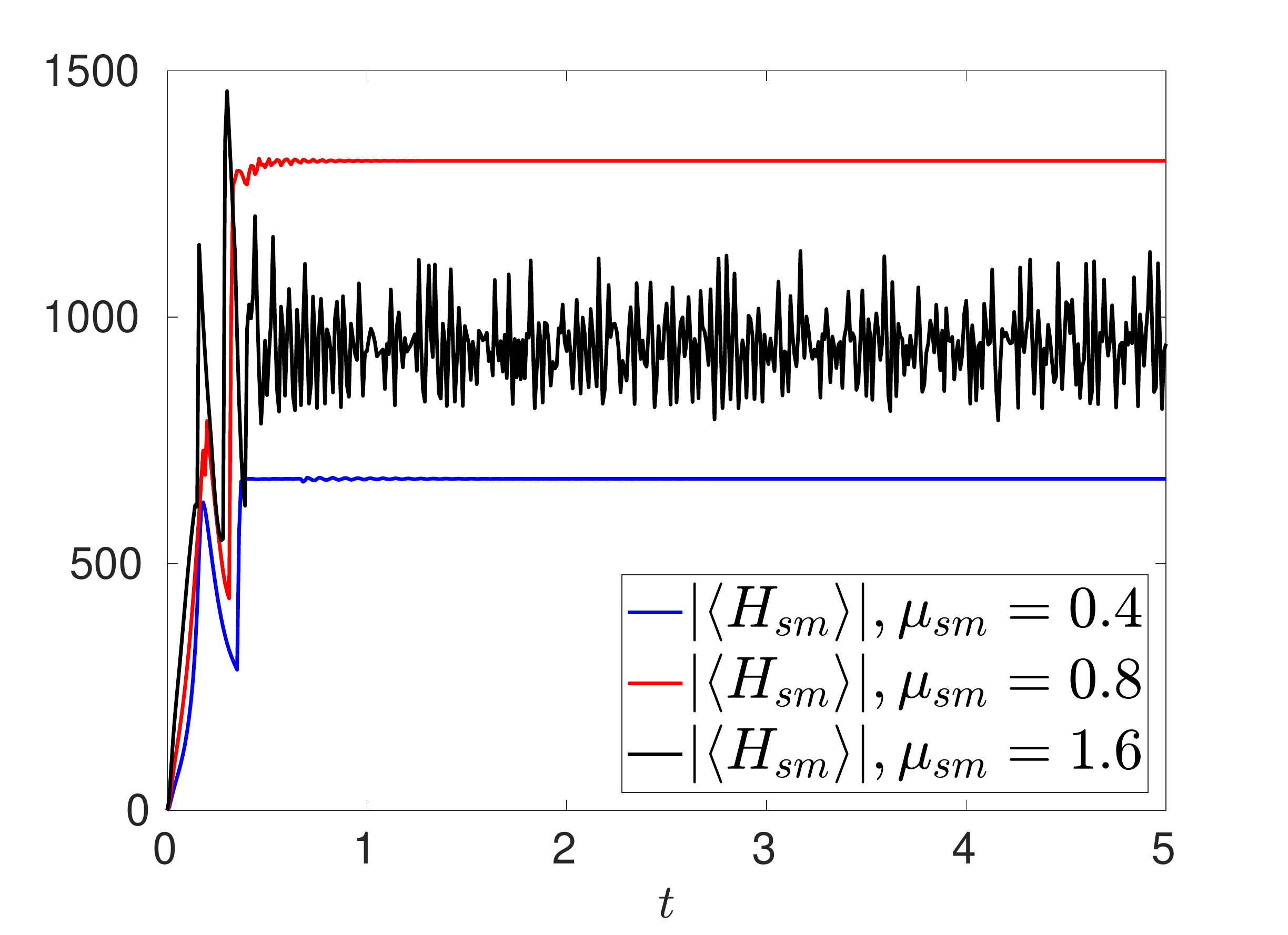}\label{Med_Hsm_comparison}}   
\caption{Number $\langle \hat N_\hf\rangle $ of healthy cells \textbf{(a)}, number $\langle \hat N_\cf\rangle $ of tumor cells \textbf{(b)},
modulus $|\langle H_{\hf\Hf}\rangle|$ of the expectation value  of the healthy cells production Hamiltonian \textbf{(c)} and modulus $|\langle H_{\cf\mf}\rangle|$ of the expectation value  of the medical treatment Hamiltonian \textbf{(d)}, 
 during the medical treatment scenario for various values of the parameter $\mu_{\cf\mf}$. Initial conditions and other parameters are those
 used in configuration $R1$}
\end{figure}
\end{center}

 \subsection{Medical treatment II: time dependent treatment }
 
We consider in this Section the possibility of introducing a time dependency in the action of the medical treatment.
In particular, we suppose that $\mu_{\cf\mf}$ can depend  both on time and on the number of the cells.

In the first configuration,
called $MT1$,  we assume that 
\bea 
\mu_{\cf\mf}(t)=\sum\limits_{k=1}^MA\exp(-\left((t-k)/\sigma\right)^2), \quad A,\sigma>0,\label{mus2}
\ena
which corresponds to a variable intensity of the medical treatment having peaks at $t=1,2,\ldots, M$. Here $M$ is the last value of $t$ when the medical treatment is acting.
Initial conditions and other parameters are as in $R1$.

\begin{center} 
\begin{figure}
\subfigure[configuration $MT1$]{\includegraphics[width=8cm]{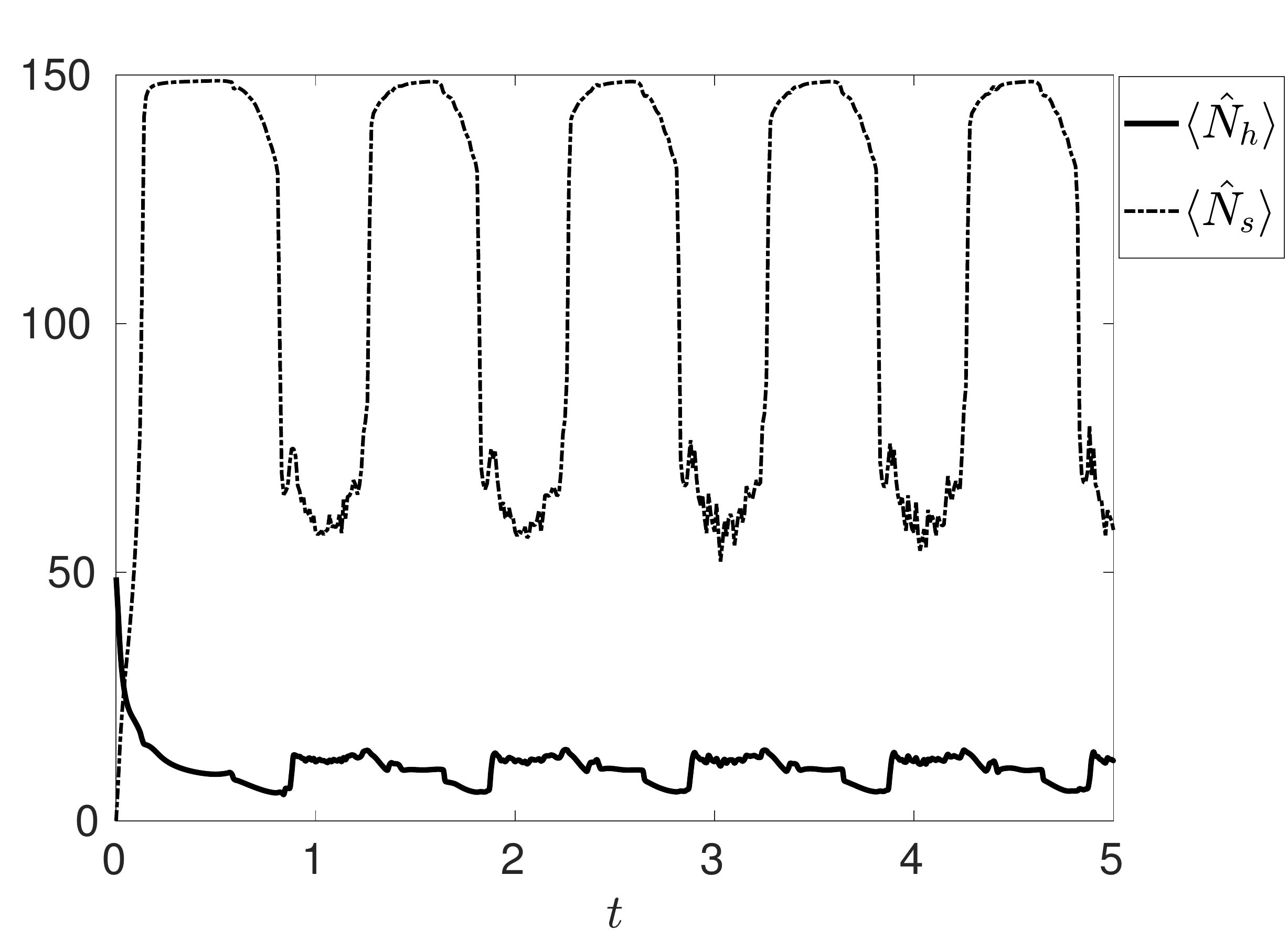}\label{Med_gaussian}}
\subfigure[configuration $MT2$]{\includegraphics[width=8cm]{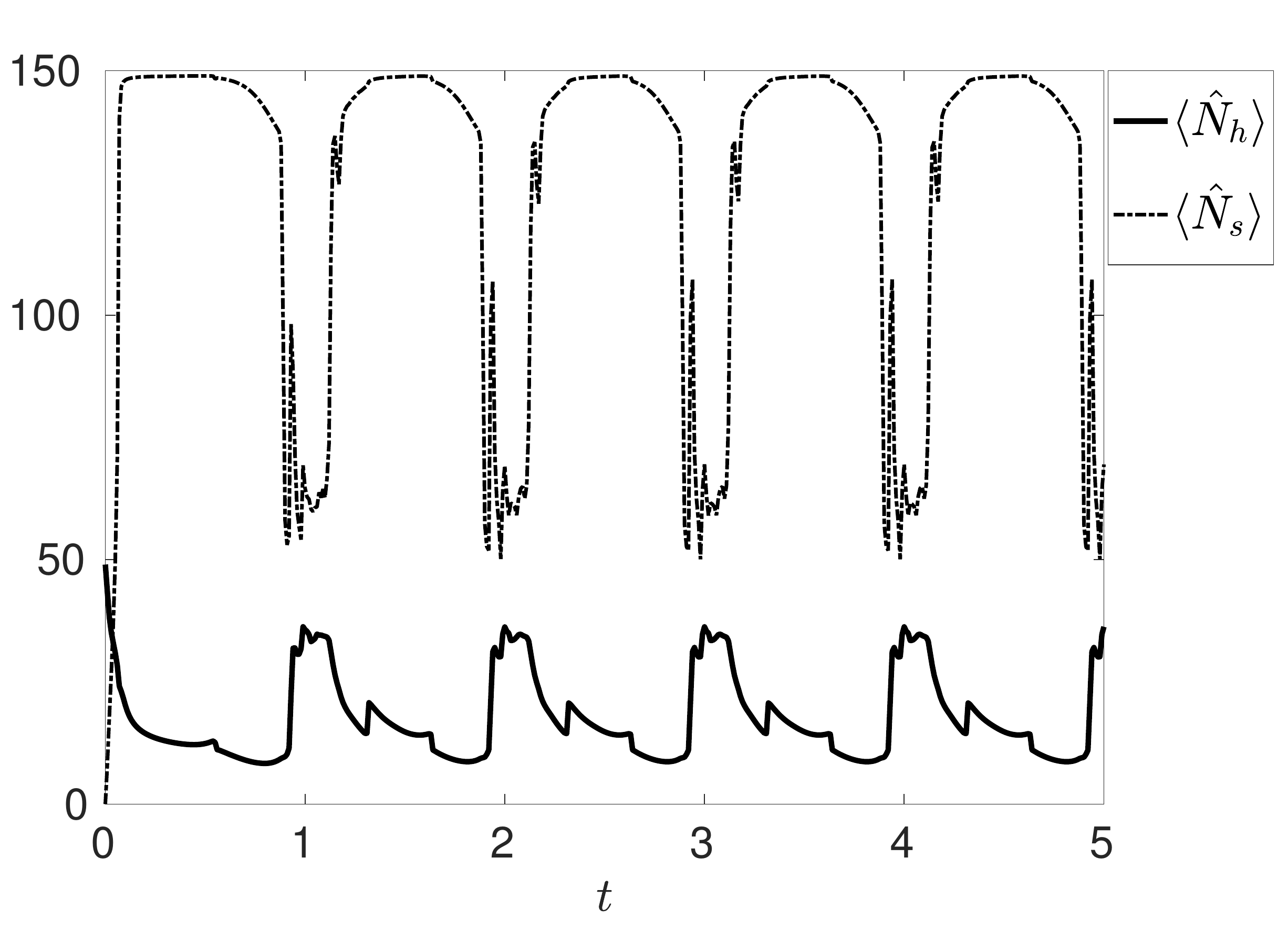}\label{Med_gaussian_sm}}
\subfigure[configuration $MT1$]{\includegraphics[width=8cm]{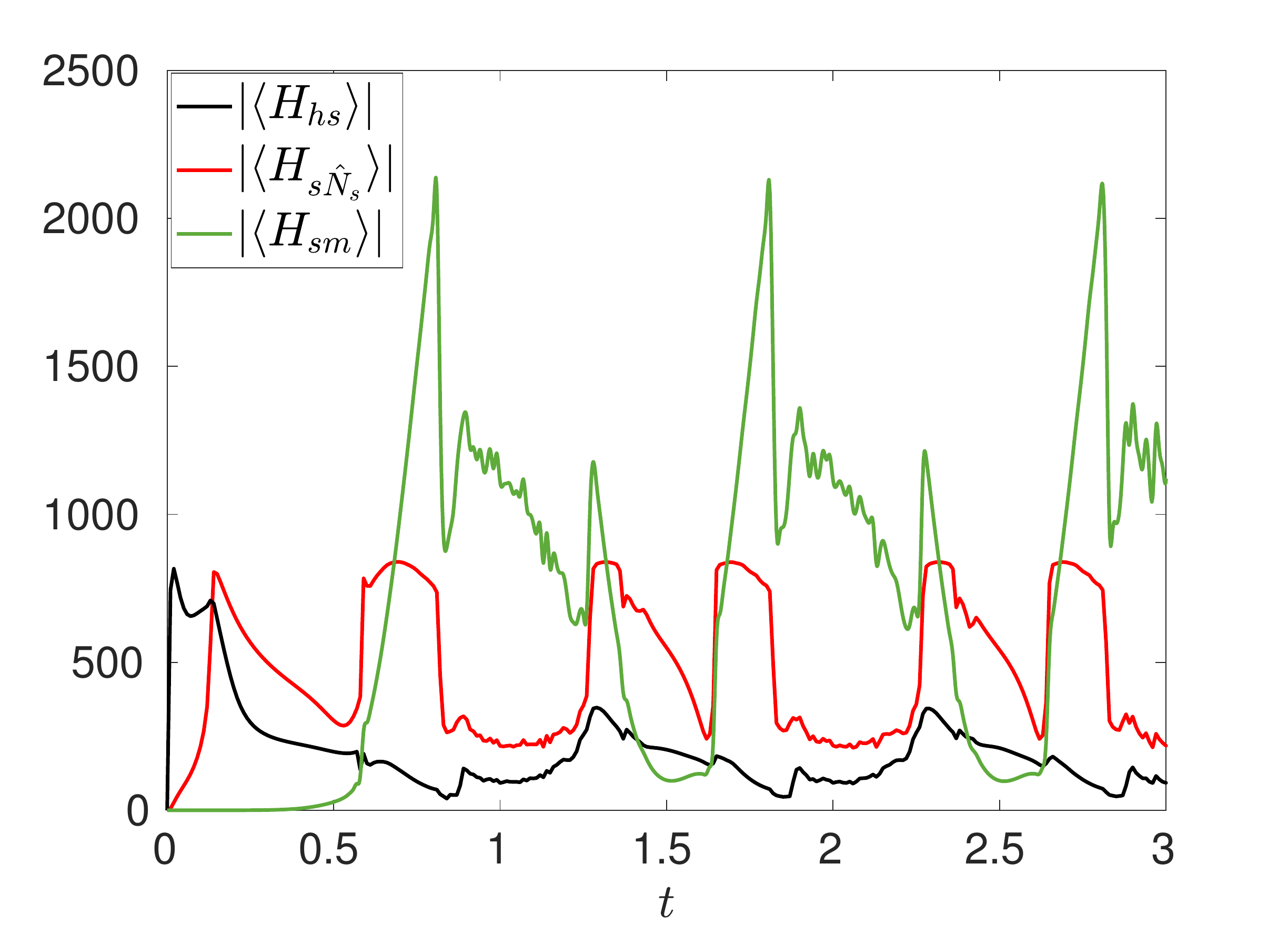}\label{Med_H_gaussian}}
\subfigure[configuration $MT2$]{\includegraphics[width=8cm]{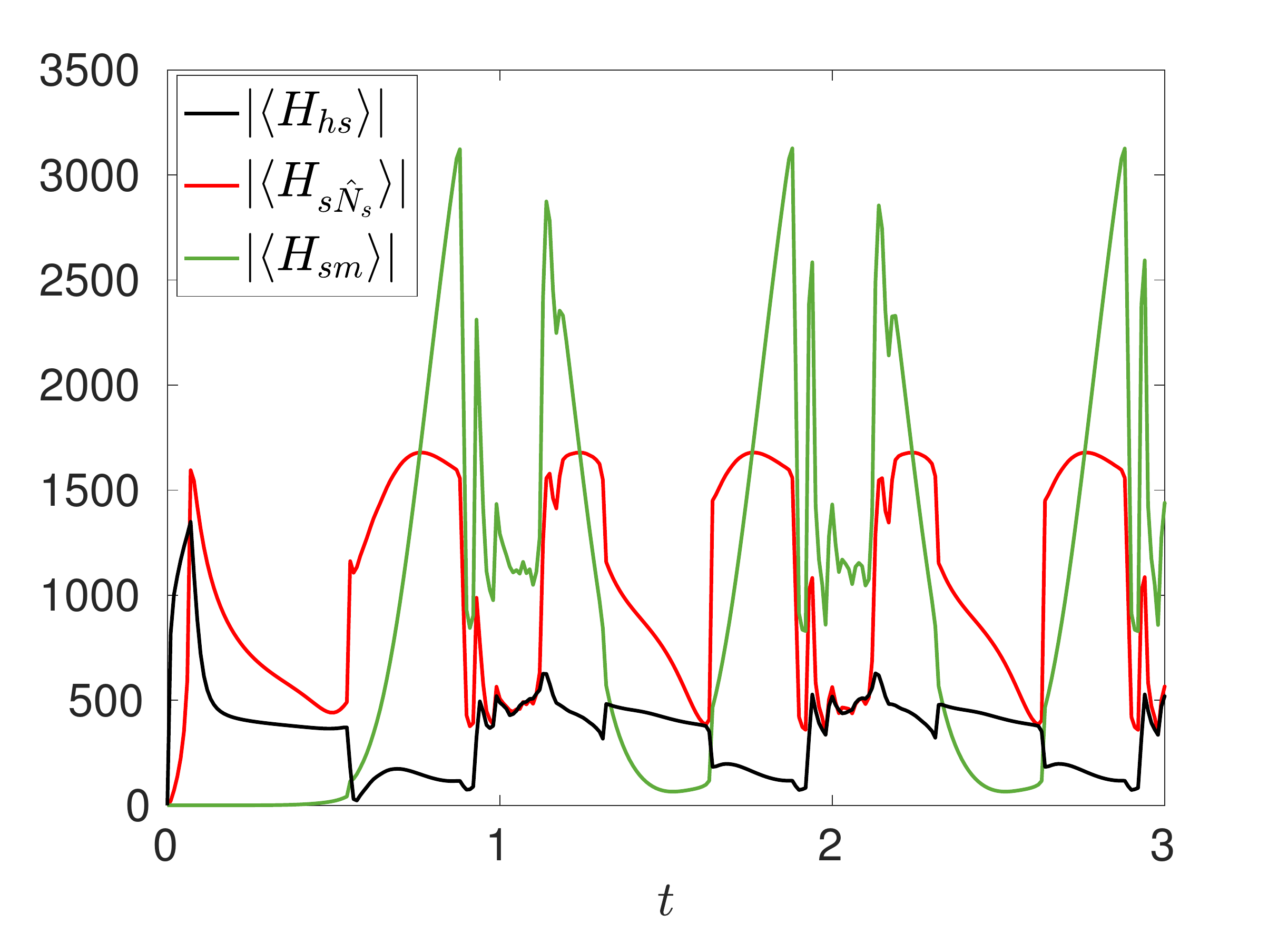}\label{Med_H_gaussian_sm}}
\caption{Numbers $\langle \hat N_\hf\rangle $ of healthy cells \textbf{(a)} and moduli  $|\langle H_{\hf\cf}\rangle|, |\langle H_{\cf\Cf}\rangle|, 
|\langle H_{\cf\mf}\rangle|$ of the 
 expectation values \textbf{(b)} for the time dependent configuration $MT1$.  Numbers $\langle \hat N_\hf\rangle $ of healthy cells \textbf{(c)} and moduli $|\langle H_{\hf\cf}\rangle|, |\langle H_{\cf\Cf}\rangle|,
 |\langle H_{\cf\mf}\rangle|$ of the 
expectation values  \textbf{(d)} for the time dependent configuration $MT2$. In both configurations, initial conditions are those of $R1$. Other parameters are specified in the text.}
\end{figure}
\end{center}

Results in  terms of number of cells, and amplitude of $|\langle H_{\hf\cf}\rangle|, |\langle H_{\cf\Cf}\rangle|,
|\langle H_{\cf\mf}\rangle|$ are shown in Figure \ref{Med_gaussian} and Figure \ref{Med_H_gaussian} for  $A=2.5,\sigma=0.25,M=4$.
During the maximum intensity of the medical treatment, at the peaks of $\mu_{\cf\mf}(t)$ and the consequent high value of $|\langle H_{\cf\mf}\rangle|$, the number of tumor cells suddenly decreases,
with a simultaneous increment of healthy cells. At the same time, the effect of the mutation of healthy into sick cells is weakened, as one can easily deduce from the plot of  $|\langle H_{\hf\cf}\rangle|$, see Figure \ref{Med_H_gaussian}.

The second configuration,  $MT2$, is based on a time-dependent choice of the parameter $\mu_{\hf\cf}$, which we assume behaves as follows:  $\mu_{\hf\cf}(t)=\frac{\tilde\mu_{\hf\cf}}{1+\mu_{\cf\mf}(t)}$, that is we assume that when the strength 
$\mu_{sm}(t)$ of the treatment increases, the mutation of healthy  into tumor cells decreases.
Moreover we set  $\tilde\mu_{\hf\cf}=1,\mu_{\cf\cf}=\mu_{\hf\hf}=1$, while the initial conditions, the other parameters and  $\mu_{\cf\mf}(t)$ are  the same specified for the configuration $MT1$.
The results are shown in Figure \ref{Med_gaussian_sm} and Figure \ref{Med_H_gaussian_sm}. The treatment here has two effects: it destroys tumor cells, and it reduces the mutation of healthy cells. Figure \ref{Med_H_gaussian_sm} shows that, when
 the  $|\langle H_{\cf\mf}\rangle|$ has its peaks, that is when the intensity of the Hamiltonian term $H_{\cf\mf}$ is higher, the number of healthy cells
significantly grows. Comparing Figures \ref{Med_gaussian} and \ref{Med_gaussian_sm} we also see that configuration $MT2$ {\em works better} than $MT1$: more tumor cells are destroyed and more healthy cells are created in correspondence of the peaks of $\mu_{\cf\mf}(t)$. However, we still do not get complete healing.

\begin{center} 
\begin{figure}
\subfigure[Configuration $MT3$]{\includegraphics[width=8cm]{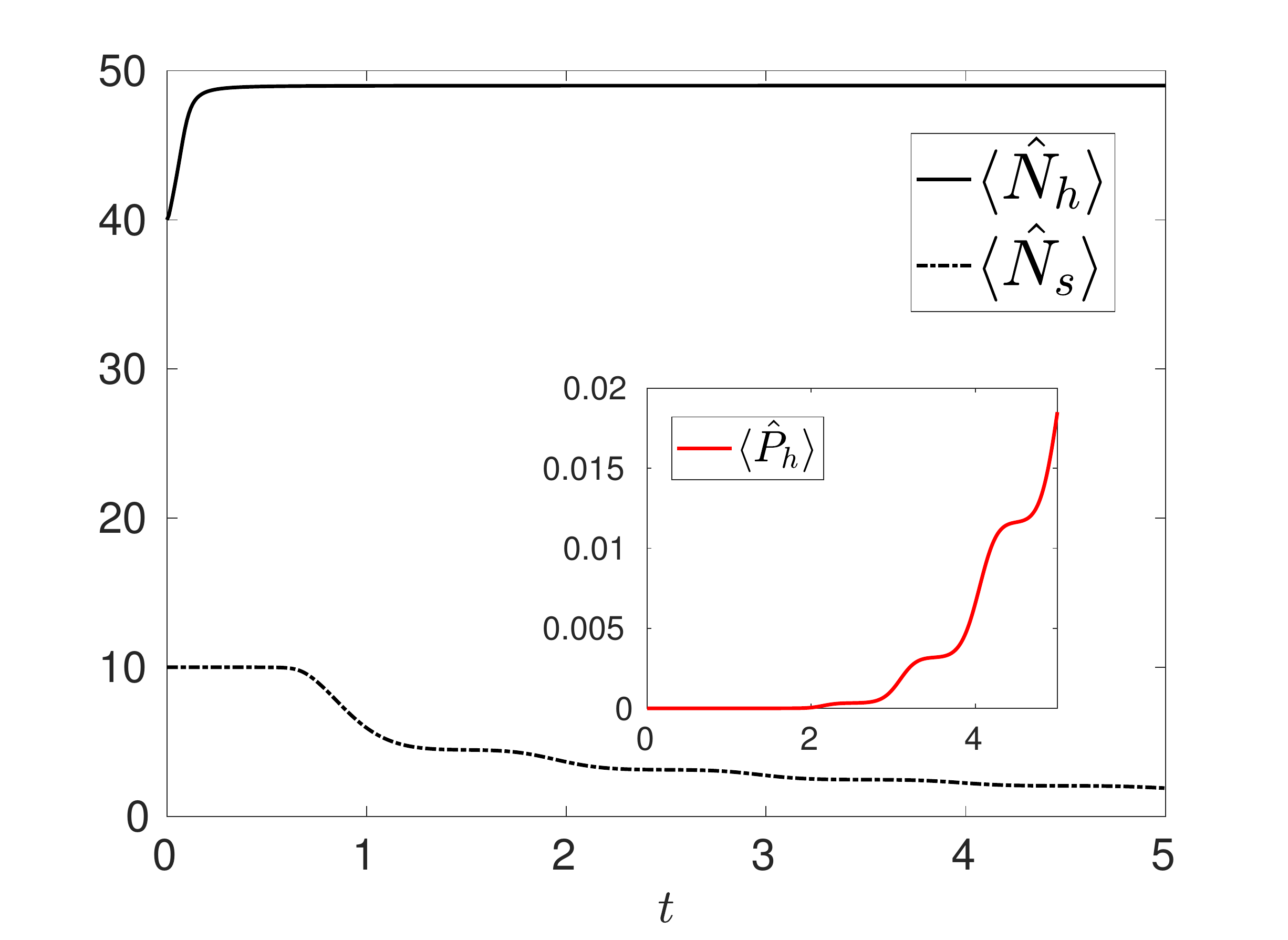}\label{Med_guarigione}}
\subfigure[Configuration $MT3$]{\includegraphics[width=8cm]{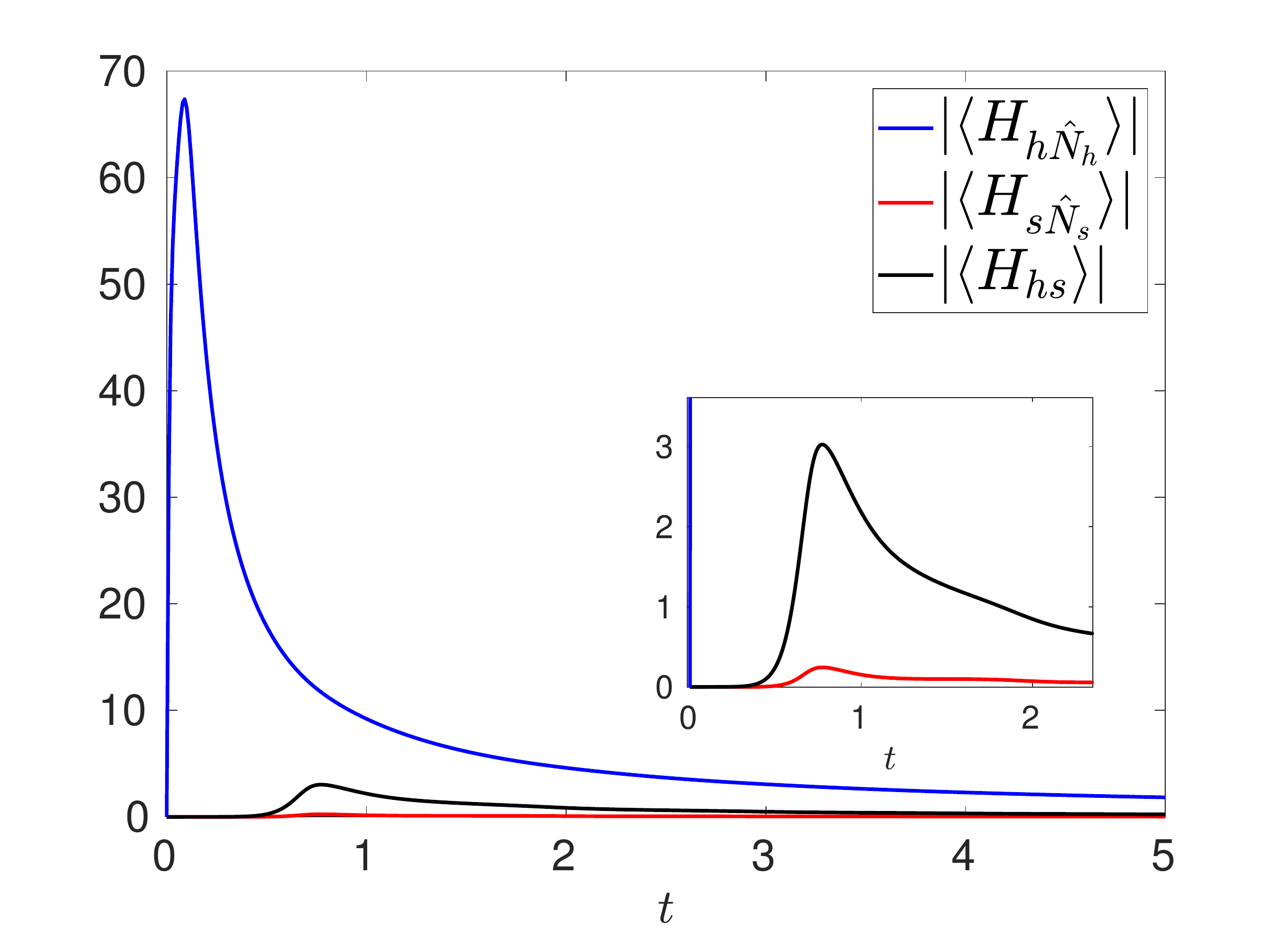}\label{Med_guarigione_H}}
\caption{Configuration $MT3$. \textbf{(a)} Numbers $\langle \hat N_\hf\rangle,\langle \hat N_\cf\rangle $ of healthy and tumor cells.
In the inset the probability $\langle \hat P_\hf\rangle$ of having only healthy cells in the system. \textbf{(b)} Time evolutions of the moduli $|\langle H_{\hf\Hf}\rangle|,|\langle H_{\hf\cf}\rangle|, |\langle H_{\cf\Cf}\rangle|$ of the expectation values. In the inset we plot the magnification of the early time evolution. 
}
\label{Med_guarigine} 
\end{figure}
\end{center}

{
For this reason, we consider a third scenario, $MT3$, in which a quasi-healthy  state is obtained, choosing properly some of the parameters of $H$.
This healing is based on the idea  of controlling and limiting the mutation of the healthy cells and the proliferation of tumor cells.
We accomplish this by considering the following time dependent parameters: 
\bea
\mu_{\hf\cf}(t)&=&\tilde \mu_{\hf\cf}\textrm{exp}\left(-\langle \Mf\rangle\frac{N_\cf-1-\langle \Cf\rangle}{\langle \Cf\rangle}\right),\label{muss1}\\
\mu_{\cf\cf}(t)&=&\tilde \mu_{\cf\cf}\textrm{exp}\left(-\langle \Mf\rangle\frac{N_\cf-1-\langle \Cf\rangle}{\langle \Cf\rangle}\right),\label{muss2}
\ena
whereas $\tilde \mu_{\hf\cf},\tilde \mu_{\cf\cf}$ are fixed constants. In \eqref{muss1}-\eqref{muss2}  $\mu_{\hf\cf}(t)$ and $\mu_{\cf\cf}(t)$ depend on time only when the medical treatment is acting ($\langle \Mf\rangle\neq 0$), and their amplitudes decreases when the number of tumor cells is low ($\mu_{\hf\cf}(t), \mu_{\cf\cf}(t)\approx0$ for $\langle \Cf\rangle\rightarrow 0$): mutation and proliferation of sick cells are small effects if $\langle \Cf\rangle$ is small. But, when $\langle \Cf\rangle$ increases, these mechanisms become more and more relevant.

We present in Figure \ref{Med_guarigione} the results of the numerical simulation of $MT3$ with initial condition 
$\Psi(0)=\varphi_{40,10,1}$, corresponding to $\langle \Hf\rangle=40,\langle \Cf\rangle=10$, $\tilde \mu_{\hf\cf}=\tilde \mu_{\cf\cf}=1$, and  $\mu_{\cf\mf}(t)$ and other parameters as in $MT1$.
The evolutions of $\langle \Hf\rangle$ and $\langle \Cf\rangle$ clearly show that the number of healthy cells increases whereas the number of tumor cells decreases, with a non vanishing probability $\langle P_\hf\rangle$ that the system contains only healthy cells.
{The main effect induced by the treatment is the overall control over the degeneracy of the healthy cells and the mitosis of the tumor cells, which are now weak mechanisms when compared to the normal proliferation of healthy cells. This can be deduced from the time evolutions of $|\langle H_{\hf\Hf}\rangle|,|\langle H_{\hf\cf}\rangle|, |\langle H_{\cf\Cf}\rangle|$, shown in Figure \ref{Med_guarigione_H}:
$|\langle H_{\hf\cf}\rangle|, |\langle H_{\cf\Cf}\rangle|$
are negligible with respect to  $|\langle H_{\hf\Hf}\rangle|$, and tend to vanish together with the number of tumor cells.}
}

\section{Conclusions}
\label{sec::conclusions}
In this paper,   we have constructed an operatorial model for mutations of healthy into tumor cells, focusing in particular on the possibility of reversing this transition. We have proposed a modified version of the general strategy proposed in \cite{bagbook}, based on a non-hermitian Hamiltonian describing, in a non reversible way, several relevant biological effect: mutation of healthy  into tumor cells, proliferation by mitosis of all the cells, and a medical treatment which acts to control and limit the proliferation of tumor cells.
The results have been presented in terms of numbers of cells and probability to have a pure healthy or a pure sick state.
Furthermore, we have  explained the various stages of the evolution of $\Sc$ in terms of  mean values of the various terms appearing in the Hamiltonian. 

We have first seen how, in absence of any medical treatment, the model describes well the effect of the carcinogenic factor causing mutation of the cells. Then, we have seen situations in which the effect of the treatment can partially reverse, or at least keep under control, this mutation. Not surprisingly, the efficiency of the treatment is related to its strength.
We have shown that, within our model,  a complete recovery of the system is possible only when both  the mutation of the healthy cells, and the mitosis of the tumor cells, are properly controlled.  


The model we have proposed, of course, is very basic, as it contains only few essential mechanisms ruling the tumor growth. Many possible changes/improvements are possible. 
For instance, one could consider a model with some spatial dependency.
Moreover a deeper analysis of the role of the starting time of the treatment is essential, in connection with early diagnosis. More {\em biologically motivated} effects should further be inserted in the Hamiltonian $H$, to make the model more realistic. For that, a comparison between our results and biological finding would be essential.

\end{document}